\documentclass[prd,aps,
twocolumn,
nofootinbib,
preprintnumbers,
superscriptaddress]{revtex4}
\usepackage[T1]{fontenc} % if needed
\usepackage{graphicx}
\usepackage{epsfig}
\usepackage{xcolor}
\usepackage{rotating}
\usepackage{amssymb}
\usepackage{subfigure}
\usepackage{dsfont}
\usepackage{psfrag}
\usepackage{amsmath,euscript,array,mathrsfs}
\usepackage{bbold}
\usepackage{epsf}
\usepackage[utf8]{inputenc}
\usepackage{placeins}
\usepackage{dcolumn}
\usepackage{tabularx}
\newcolumntype{Y}{>{\centering\arraybackslash}X}

\newcommand{\beq}{\begin{equation}}
\newcommand{\eeq}{\end{equation}}
\newcommand{\beqs}{\begin{eqnarray}}
\newcommand{\eeqs}{\end{eqnarray}}

\newcommand{\gsim}{\mathrel{\raisebox{-.6ex}{$\stackrel{\textstyle>}{\sim}$}}}
\newcommand{\Tr}{{\rm Tr}}

\def\hbar{\hspace{0pt}\raisebox{1pt}{$-$} \hspace{-7pt} h}

\newcommand{\be}{\begin{equation}}
\newcommand{\ee}{\end{equation}}

\newcommand{\bea}{\begin{eqnarray}}
\newcommand{\eea}{\end{eqnarray}}

\def\lbldef#1#2{\expandafter\gdef\csname #1\endcsname {#2}}

\def\href#1#2{#2}

\newcommand{\ber}{\begin{eqnarray}}
\newcommand{\eer}{\end{eqnarray}}

\newcommand{\beqar}{\begin{eqnarray}}

\newcommand{\eeqar}{\end{eqnarray}}

\newcommand{\dsl}
  {\kern.06em\hbox{\raise.15ex\hbox{$/$}\kern-.56em\hbox{$\partial$}}}

\newcommand{\eeqarr}{\end{eqnarray}}
\newcommand{\ZZ}{{\rm \kern 0.275em Z \kern -0.92em Z}\;}

\def\CC{{\mathchoice
{\rm C\mkern-8mu\vrule height1.45ex depth-.05ex
width.05em\mkern9mu\kern-.05em}
{\rm C\mkern-8mu\vrule height1.45ex depth-.05ex
width.05em\mkern9mu\kern-.05em}
{\rm C\mkern-8mu\vrule height1ex depth-.07ex
width.035em\mkern9mu\kern-.035em}
{\rm C\mkern-8mu\vrule height.65ex depth-.1ex
width.025em\mkern8mu\kern-.025em}}}

\def\RR{{\rm I\kern-1.6pt {\rm R}}}

\def\ZZ{{\rm Z}\kern-3.8pt {\rm Z} \kern2pt}
\def\IB{\relax{\rm I\kern-.18em B}}
\def\ID{\relax{\rm I\kern-.18em D}}
\def\II{\relax{\rm I\kern-.18em I}}
\def\IP{\relax{\rm I\kern-.18em P}}

\newcommand{\bear}{\begin{eqnarray}}
\newcommand{\eear}{\end{eqnarray}}

\def\to{\rightarrow}

\def\to{\rightarrow}

\def\i{\iota}

\def\6{\partial}

\def\bea{\begin{eqnarray}}
\def\eea{\end{eqnarray}}

\def\beqx{\begin{displaymath}}
\def\eeqx{\end{displaymath}}

\newcommand{\bmat}{\left(\begin{array}}
\newcommand{\emat}{\end{array}\right)}

\def\i{\iota}

\def\bo{{\raise-.3ex\hbox{\large$\Box$}}}               % D'Alembertian
\def\face{{\raise.2ex\hbox{$\displaystyle \bigodot$}\mskip-2.2mu \llap {$\ddot \smile$}}}% happy face
\def\>{\rangle}                                      %right angle
\def\<{\langle}                                      %left angle

\def\leftrightarrowfill{$\mathsurround=0pt \mathord\leftarrow \mkern-6mu
        \cleaders\hbox{$\mkern-2mu \mathord- \mkern-2mu$}\hfill
        \mkern-6mu \mathord\rightarrow$}        % <--> double differential
\def\dvec#1{\vbox{\ialign{##\crcr
        \leftrightarrowfill\crcr\noalign{\kern-1pt\nointerlineskip}
        $\hfil\displaystyle{#1}\hfil$\crcr}}}           % <--> accent
\def\Tr{{\rm Tr \,}}                                    % Trace
\def\-{\hphantom{-}}

\begin{document}
\title{$Sp(2N)$ Yang-Mills theories on the lattice: scale setting and topology}

\author{Ed Bennett}
\email{e.j.bennett@swansea.ac.uk}
\affiliation{Swansea Academy of Advanced Computing, Swansea University, Fabian Way, SA1 8EN, Swansea, Wales, UK}
\author{Deog Ki Hong}
\email{dkhong@pusan.ac.kr}
\affiliation{Department of Physics, Pusan National University, Busan 46241, Korea}
\author{Jong-Wan Lee}
\email{jwlee823@pusan.ac.kr}
\affiliation{Department of Physics, Pusan National University, Busan 46241, Korea}
\affiliation{Institute for Extreme Physics, Pusan National University, Busan 46241, Korea}
\author{C.-J.~David~Lin}
\email{dlin@nycu.edu.tw}
\affiliation{Institute of Physics, National Yang Ming Chiao Tung University, 1001 Ta-Hsueh Road, Hsinchu 30010, Taiwan}
\affiliation{Center for High Energy Physics, Chung-Yuan Christian University,
Chung-Li 32023, Taiwan}
\affiliation{Centre for Theoretical and Computational Physics, National Yang Ming Chiao Tung University, 1001 Ta-Hsueh Road, Hsinchu 30010, Taiwan}
\affiliation{Physics Division, National Centre for Theoretical Sciences, Taipei 10617, Taiwan}
\author{Biagio Lucini}
\email{b.lucini@swansea.ac.uk}
\affiliation{Department of Mathematics, Faculty  of Science and Engineering,
Swansea University, Fabian Way, SA1 8EN Swansea, Wales, UK}
\affiliation{Swansea Academy of Advanced Computing, Swansea University,
Fabian Way, SA1 8EN, Swansea, Wales, UK}
\author{Maurizio Piai}
\email{m.piai@swansea.ac.uk}
\affiliation{Department of Physics, Faculty  of Science and Engineering,
Swansea University,
Singleton Park, SA2 8PP, Swansea, Wales, UK}
\author{Davide Vadacchino}
\email{davide.vadacchino@plymouth.ac.uk}
\affiliation{School of Mathematics and Hamilton Mathematics Institute, Trinity
College, Dublin 2, Ireland}
\affiliation{Centre for Mathematical Sciences, University of Plymouth, Plymouth, PL4 8AA, United Kingdom}

%\date{\today}

\begin{abstract}
We study
Yang-Mills lattice theories with $Sp(N_c)$ gauge group, with
$N_c=2N$, for
$N=1,\,\cdots,\,4$.
We show that if we divide  the renormalised
couplings appearing in the Wilson flow by
 the quadratic Casimir $C_2(F)$ of the $Sp(N_c)$ group,
then the resulting quantities
display a good agreement among all values of $N_c$ considered, 
over a finite interval in flow time.
We use this scaled version of the Wilson flow
as a scale-setting procedure,
compute the topological susceptibility of 
the $Sp(N_c)$ theories, and
extrapolate the results to the continuum limit for each $N_c$.
\end{abstract}

\maketitle
\tableofcontents
\preprint{PNUTP-22/A03}

%%%%%%%%%%%%%%%%%%%%%%%%%%%%%%%%%%%
\section{Introduction}

Lattice studies  of $Sp(N_c=2N)$ gauge theories
aim at quantitatively appraising, in the strong coupling regime, 
what are their distinctive features in respect to theories
 based upon $SU(N_c)$ gauge groups.
For example, a long list of recent investigations of the 
 $SU(2)\sim Sp(2)$ theories~\cite{Hietanen:2014xca,Detmold:2014kba,
 Arthur:2016dir, Arthur:2016ozw, Pica:2016zst,Lee:2017uvl,Drach:2017btk,
 Drach:2020wux,Drach:2021uhl},
and  of the $Sp(2N)$ theories for $N>1$~\cite{Bennett:2017kga,
 Lee:2018ztv,Bennett:2019jzz,Bennett:2019cxd,Bennett:2020hqd,Bennett:2020qtj,
 Bennett:2022yfa,AS}, are  motivated by 
their paradigm changing potential for applications
in contemporary high energy physics.

Prominently,  enhanced global symmetry and symmetry breaking patterns
arise in $Sp(2N)$ theories in the presence of matter fields.
This feature is exploited in new-physics  model-building exercises 
such as in the minimal model in
 Ref.~\cite{Barnard:2013zea}, which  combines a
 composite Higgs model (CHM)--- in which the Higgs fields originate as  
 pseudo-Nambu-Goldstone bosons (PNGBs)
in a new strongly coupled sector~\cite{Kaplan:1983fs,Georgi:1984af,Dugan:1984hq}---with 
partial top compositeness~\cite{Kaplan:1991dc}. 
 For recent reviews see Refs.~\cite{Panico:2015jxa,
 Witzel:2019jbe,Cacciapaglia:2020kgq}, and the summary tables in 
 Refs.~\cite{Ferretti:2013kya,Ferretti:2016upr,Cacciapaglia:2019bqz}.
In the different context of dark matter models emerging from 
 strongly-coupled dynamics~\cite{Hochberg:2014dra,Hochberg:2014kqa,Hochberg:2015vrg,Kondo:2022lgg},
$Sp(2N)$ gauge theories have also been attracting increasing interest~\cite{Bernal:2017mqb,
 Berlin:2018tvf,Bernal:2019uqr,Cai:2020njb,Tsai:2020vpi,Maas:2021gbf,Zierler:2021cfa,Kulkarni:2022bvh}.

There are more general, theoretical reasons to
 study $Sp(2N)$ gauge theories.
Pioneering studies of the pure gauge cases~\cite{Holland:2003kg}
aimed at qualifying the role of
 the centre symmetry in  the confinement/deconfinement transition at
finite temperature.
In the light of the conjectured existence of dualities between large-$N_c$
gauge theories and theories of gravity in higher 
dimension~\cite{Maldacena:1997re,Gubser:1998bc,Witten:1998qj,Aharony:1999ti},
the appeal of $Sp(N_c)$ theories 
derives  from the common features that they share
 with $SU(N_c)$ theories.
For example, extensive studies of the spectrum of glueballs
and strings confirm that universal features emerge at large $N_c$,
common to $SU(N_c)$, $Sp(N_c)$, and $SO(N_c)$ theories~\cite{Lucini:2001ej,
Lucini:2004my, Lucini:2010nv, Lucini:2012gg, 
Athenodorou:2015nba,
Lau:2017aom,  Hong:2017suj, Yamanaka:2021xqh, 
Hernandez:2020tbc, Athenodorou:2021qvs, Bonanno:2022yjr,
Bennett:2020hqd,
Bennett:2020qtj}.

 The topological susceptibility, $\chi$
(to be defined in the body of the paper),
plays a central role in our understanding of QCD,
 for several intercorrelated reasons of historical significance.
It enters the Witten-Veneziano formula~\cite{Witten:1979vv,Veneziano:1979ec}
for the large-$N_c$ behaviour of the mass of the $\eta^{\prime}$ meson,
and the solution to the $U(1)_A$ problem.
Allowing the gauge coupling to be complex, 
 and expanding in powers of (small) $\theta$,  $\chi$
 appears as the coefficient at $O(\theta^2)$ in the free energy.
Indirectly, it might hence
have implications for the strong-CP problem, for the physics of putative 
new particles such as the axion,
and for the electric dipole moments of hadrons.
Many interesting lattice calculations of $\chi$  exist---see
Refs.~\cite{Lucini:2001ej,DelDebbio:2002xa,
Lucini:2004yh,DelDebbio:2004ns,Luscher:2010ik,Panagopoulos:2011rb,
Bonati:2015sqt,Bonanno:2020hht,
Ce:2016awn,Bonati:2016tvi,Alexandrou:2017hqw,Borsanyi:2021gqg,Cossu:2021bgn,
Athenodorou:2021qvs,Teper:2022mmj}, and
Tables~1 and~2 of  the review in Ref.~\cite{Vicari:2008jw}.

Accounting for $\chi$
might provide  new insight in the role  of instantons
and other non-perturbative objects.
A precise knowledge of $\chi$ might  have unexpectedly important consequences also
in the aforementioned  subfield encompassing  modern phenomenological
applications of strongly-coupled $Sp(N_c)$ theories---models of composite Higgs, partial top
compositeness, dark matter, or even early universe physics.
In general, precise calculations of $\chi$ might
sharpen our understanding both of  commonalities and  differences between 
$Sp(N_c)$ and $SU(N_c)$ theories,
starting in the Yang-Mills (pure gauge) theories.

Motivated by such considerations, and as an important step in the programme 
of study  of $Sp(N_c)$ lattice gauge theories conducted 
by our collaboration, in this paper we 
compute the topological susceptibility of $Sp(N_c)$ Yang-Mills theories for $N_c\leq 8$.
We made available  preliminary results  in contributions to Conference
 Proceedings~\cite{Lucini:2021xke,Bennett:2021mbw}, but this analysis 
is much improved,
and based on  larger statistics.
In a dedicated publication, we  compare our results for $Sp(N_c)$ with the $SU(N_c)$
literature, and discuss the large-$N_c$ extrapolation~\cite{Bennett:2022gdz}.

The paper is organised as follows.
In Sect.~\ref{Sec:theory} we define the lattice theories of interest.
In Sect.~\ref{Sec:scale_setting} we discuss how to 
use the gradient flow and its lattice
 implementation, the Wilson flow~\cite{Luscher:2010iy,Luscher:2013vga},
 as a scale setting procedure, and
define  the topological charge and susceptibility.
Sect.~\ref{Sec:numerical} is the main body of the paper,
in which we present our numerical results.
We conclude with the summary in Sect.~\ref{Sec:outlook}.
We relegate some useful details to Appendix~\ref{Appendix}.

%%%%%%%
%%%%%%%
\section{$Sp(2N)$ Yang-Mills theories}
\label{Sec:theory}

We  define the 
continuum   $Sp(2N)$ gauge theories, in $4$-dimensional Euclidean space,
in terms of  the action
\beq\label{eq:YMaction}
S_\mathrm{YM} \equiv -\frac{1}{2g_0^2} \int\mathrm{d}^4x\,
\Tr F_{\mu\nu}F^{\mu\nu}~,
\eeq
where $g_0$ is the  gauge coupling, $F_{\mu\nu} \equiv \sum_A F_{\mu\nu}^A \tau^A$, with
\beq
F_{\mu\nu}^A = \partial_\mu A_\nu^A - \partial_\nu A_\mu^A
+ f^{ABC}A_\mu^B A_\nu^C\,,
\eeq
and the trace is over the color index on the fundamental representation, while 
 $A=1,\cdots,N(2N+1)$. 
The Hermitian matrices $\tau^A\in\mathbb{C}^{2N\times 2N}$ are the generators
of the algebra associated to the Lie group $Sp(2N)$ in the
fundamental representation. They satisfy the relations
\beq
\left[ \tau^A,\, \tau^B\right] = \dot{\imath} f^{ABC} \tau^C\,,
\eeq
and are normalised according to $\Tr\left( \tau^A \tau^B\right)=\frac{1}{2}\delta^{AB}$.

The configuration space of this theory can be partitioned into sectors, characterised by the value of
the \emph{topological charge} $Q$, defined as follows:
\begin{equation}
    Q \equiv \int \mathrm{d}^4 x~q(x)\,,
\end{equation}
where
\begin{equation}
\label{Eq:q}
    q(x) \equiv \frac{1}{32\pi^2}
    \epsilon^{\mu\nu\rho\sigma}\,
    \mathrm{Tr}~F_{\mu\nu}(x) F_{\rho\sigma}(x)\,.
\end{equation}
The \emph{topological susceptibility}, $\chi$, 
is defined as
\begin{equation}
    \chi \equiv \int \mathrm{d}^4 x ~\langle q(x) q(0) \rangle.
\end{equation}
The possible values of $Q$ belong to the third homotopy group of the gauge group. Since
$Sp(2N)$ is compact, connected and simple, one finds that
\begin{equation}
    \pi_3( Sp(2N) ) = \mathbb{Z}\,,
\end{equation}
as in the case of $SU(N_c)$ gauge theories.

%%%%%%%%%%%%%%%%%
%%%%%%%%%%%%%%%%%
\subsection{The lattice}
\label{Sec:lattice}

A lattice regularisation of the theory defined in Eq.~(\ref{eq:YMaction})
allows to characterise quantitatively its non-perturbative features.
We adopt a $4$-dimensional
Euclidean hypercubic lattice $\Lambda$, with lattice spacing $a$. The
sites of the lattice are denoted by their Cartesian 
coordinates $x=\left\{x_\mu\right\}$ and the links by $(x,\,\mu)$, where $\mu=0,\,\cdots,\,3$ 
and $x_\mu = n_\mu a$. 
For a lattice of length $L_\mu $ in the $\mu$ directions, with $n_\mu=0,\,\cdots,\,L_\mu/a-1$, 
 the total number of sites is thus $V_4/a^4= L_0 L_1 L_2 L_3/a^4 $. 
 The lattices used in our calculations are isotropic in the four directions, $L_{\mu}=L$, 
and we impose periodic boundary conditions in all directions. The elementary degrees of freedom of the theory are
called \emph{link variables}, and defined as
\beq
U_\mu(x) \equiv \exp\left( \dot{\imath} \int_x^{x+\hat{\mu}}\mathrm{d} \lambda^\mu \tau^A 
A_\mu^A(\lambda) \right)\,,
\eeq
where $\hat{\mu}$ is the unit vector in direction $\mu$. The link variables are $2N\times2N$ matrices that, 
under the action of a gauge transformation  $g(x) \in Sp(2N)$, transform as 
\beq
U_\mu(x) \to g(x) U_\mu(x) g^\dag(x+\hat{\mu})\,.
\eeq
The trace of a path-ordered product of link variables defined along  a closed lattice path is hence  gauge invariant.

The simplest such closed path  on the lattice
defines the  \emph{elementary plaquette} $\mathcal{P}_{\mu\nu}$:
\beq\label{eq:elementary_plaquette}
\mathcal{P}_{\mu\nu}(x) \equiv U_\mu(x) U_\nu(x+\hat{\mu})
U^\dag_\mu(x+\hat{\nu}) U^\dag_\nu(x)\,,
\eeq
and  is used to define 
the
\emph{Wilson action} $S_\mathrm{W}$ of the lattice gauge theory (LGT):
\beq\label{eq:Wilson_action}
S_\mathrm{W} \equiv \beta \sum_x \sum_{\mu<\nu} \left( 1-
\frac{1}{2N}\Re \Tr \mathcal{P}_{\mu\nu} \right)\,,
\eeq
where 
the \emph{inverse coupling} $\beta$ is defined as
\beq
\beta \equiv \frac{4N}{g_0^2}\,.
\eeq

Another operator that is useful in lattice calculations is the \emph{clover-leaf plaquette} operator, 
defined as~\cite{Sheikholeslami:1985ij,Hasenbusch:2002ai}
\begin{widetext}
\beqs
\label{Eq:clover}
\mathcal{C}_{\mu\nu}(x) &\equiv& \frac{1}{8}
\left\{ \frac{}{}U_\mu(x) U_\nu(x+\hat{\mu}) U^\dag_\mu(x+\hat{\nu})U_\nu^\dag(x)         + U_\nu(x) U^\dag_\mu(x+\hat{\nu}-\hat{\mu}) U^\dag_\nu(x-\hat{\mu})U_\mu(x-\hat{\mu})\,+\right.\\
        &&\left.
        + \,U^\dag_\mu(x-\hat{\mu}) U^\dag_\nu(x-\hat{\nu}-\hat{\mu}) U_\mu(x-\hat{\nu}-\hat{\mu})
        U_\nu(x-\hat{\nu})  
              + U^\dag_\nu(x-\hat{\nu}) U_\mu(x-\hat{\nu}) U_\mu(x-\hat{\nu}+\hat{\mu})U_\mu^\dag(x)
        -\textrm{h.c.}\frac{}{}\right\}\,.\nonumber
        \eeqs
\end{widetext}
This operator is used in the literature as a way to improve the Yang-Mills lattice action, particularly 
in the presence of fermions. In the context of this paper, it serves two purposes: we use it to test 
the regularisation dependence of our scale-setting procedure, but also in the definition of the lattice
counterparts of $Q$ and $\chi$.

Vacuum expectation values of operators $\mathcal{O}(U_\mu)$ 
built of link variables are formally defined as ensemble averages:
\beq
\label{eq:eq:vac_exp_val}
\langle {\cal O} \rangle \equiv
\frac{ \int \mathcal{D} U_\mu e^{-S_\mathrm{W}} \mathcal{O}(U_\mu)}{Z(\beta)}~,
\eeq
where $\mathcal{D}U_\mu \equiv \prod_{x,\mu} \mathrm{d}U_\mu(x)$, 
$\mathrm{d}U_\mu(x)$ being the Haar measure on $Sp(2N)$, while
\beq\label{eq:partition_function}
Z(\beta) \equiv \int \mathcal{D} U_\mu ~e^{-S_\mathrm{W}}\,
\eeq
is the partition function of the system.

For a given value of
$\beta$ and $L/a$, 
ensembles  are generated by  a Markovian
process that updates the values of the link variables in a configuration. 
The update algorithm  must respect detailed balance 
and have
equilibrium distribution $e^{-S_\mathrm{W}}$. 
An update of all the links of the lattice is called a \emph{lattice sweep}. 
It is customary to repeat the update process,  
subsequent configurations $i$ and $i+1$ in the ensemble
being separated by a fixed number $N_{\rm sw}$ of  sweeps.
The ensemble
average takes the simpler form
\beq\label{eq:ens_avg}
\langle \mathcal{O} \rangle = \lim_{M\to\infty} 
\sum_{i=1}^M \mathcal{O}_i\,,
\eeq
with $\mathcal{O}_i$ the observable $\mathcal{O}$ evaluated
on configuration $i$. 
The algorithm we adopt  combines
local heat bath (HB) and over-relaxation (OR) updates,
implemented in an openly-available \cite{hirep-repo} adaptation of
the \textrm{HiRep} code~\cite{DelDebbio:2008zf}
 to  $Sp(2N)$ groups~\cite{Bennett:2020qtj}.

    The discretised topological charge density
can be defined in several different ways~\cite{Campostrini:1989dh,Alexandrou:2017hqw},
that  differ by terms proportional to a power of $a$.
For the body of this paper, we use the clover-leaf discretisation, 
\begin{equation}
\label{Eq:qL}
    q_L(x) \equiv \frac{1}{32 \pi^2}
    \epsilon^{\mu\nu\rho\sigma}
    \mathrm{Tr}~\mathcal{C}_{\mu\nu}(x) \mathcal{C}_{\rho\sigma}(x)~.
\end{equation}
 Both  clover-leaf and elementary plaquette definitions of $q_L(x)$---the latter obtained by replacing
$\mathcal{C}_{\mu\nu}(x)$ with $\mathcal{P}_{\mu\nu}$ in Eq.~(\ref{Eq:qL})---converge 
to $q(x)$ in Eq.~(\ref{Eq:q}), 
 as $a\to0$.  But the clover-leaf definition
 treats all lattice directions symmetrically.
The (lattice) topological charge is thus
\begin{equation}
    Q_L = \sum_x q_L(x)\,,
\end{equation}
and its susceptibility is
\begin{equation}
    \chi_L = \sum_x \langle q_L(x) q_L(0) \rangle \,.
\end{equation}

Estimates of physical quantities obtained 
for given values of $\beta$ and $L/a$ 
are affected by several types of 
systematic errors. Finite size (or volume)
effects arise when probing the system over physical distances 
that are not much smaller than $L $. This systematic error becomes insignificant
if an 
increase in $L/a$ has an effect that is smaller than statistical 
fluctuations. Studies of the 
topology in $SU(N_c)$ gauge theories 
show that finite size effects are negligible provided 
$\sqrt{\sigma} L \gsim 3$, where $\sigma$ is the string tension---see, e.g., 
Figs.~3 and~4 of Ref.~\cite{Bonati:2016tvi}.
We  use  earlier analysis of
the $Sp(N_c)$ spectrum~\cite{Bennett:2020qtj} to identify
 regions of  parameter space satisfying this condition.

The evaluation of $\chi$ via lattice methods is particularly
challenging, affected by specific 
systematic effects. First, the configuration space of
the lattice theory is simply connected. Topological
sectors, and discrete topological charges, are recovered only in the 
vicinity of the continuum limit~\cite{Luscher:1981zq},
while $Q_L$ is not integer,
which affects $\chi_L$.

Second,  it is challenging to control the continuum extrapolation.
$\chi_L$ is particularly sensitive to discretisation effects,
quantum UV fluctuations  yielding
 both additive and multiplicative renormalisation~\cite{Campostrini:1989dh,Alexandrou:2017hqw,Vicari:2008jw}.
 We
extract $\chi_L$ from Wilson-flowed configurations, as we describe in Sec.~\ref{Sec:scale_setting},
hence adopting a scale-setting procedure that is also  used to smoothen out  such divergences.

 Third, 
the evaluation of $\chi_L$ in $SU(N_c)$ theories is
hindered by the divergence of the integrated autocorrelation time $\tau_Q N_{sw}$, as $a\to 0$~\cite{DelDebbio:2002xa}. 
This phenomenon, known as \emph{topological freezing},
descends from the intrinsic  difficulty of evolving 
with a  local update algorithm
a global property such as the 
topological charge. 
We expect the same challenge to arise in $Sp(2N)$ theories. 
Several ideas have been
put forward to overcome topological 
freezing~\cite{Luscher:2011kk,Cossu:2021bgn,Bonanno:2020hht,Borsanyi:2021gqg,Endres:2015yca,Luscher:2017cjh}, 
but we defer
their use to future high precision studies. Here, 
we limit ourselves to monitoring the values of $\tau_Q$, 
and discarding compromised ensembles.

%%%%%%%%%%%%%%%%%
%%%%%%%%%%%%%%%%%
\section{Scale setting and Topology}
\label{Sec:scale_setting}

The definition of the continuum limit
requires 
the implementation of a scale-setting procedure.
A scale is introduced by selecting a dimensional quantity that can (in principle) be measured 
both in the physical limit and on the lattice. 
All physical quantities  $\langle {\cal O} \rangle$ are  expressed
in terms of such scale, and measurements are repeated by varying 
the lattice parameters (in the present case, $\beta$).
The extrapolation towards $a\to 0$  yields then a finite
 value of   $\langle {\cal O} \rangle$, in the chosen units.

The string tension, $\sigma$, of Yang-Mills theories is defined as the coefficient of the 
linear term of the potential between an infinitely massive,  \emph{static} pair of
fermion and anti-fermion transforming in the fundamental representation,
in the regime of  asymptotically large separation.  On the lattice, it can be extracted
 from the asymptotic behaviour of appropriately defined 2-points correlators 
in Euclidean time. 
Thanks to its direct physical interpretation, 
$\sigma$ is often used for scale-setting in studies 
of the properties of the confining phase of pure gauge 
theories---see, e.g., 
Refs.~\cite{Lucini:2001ej,DelDebbio:2002xa,Bonati:2016tvi,Bonanno:2020hht,Athenodorou:2021qvs,Bennett:2022gdz}.
However, it suffers from the effect of both systematic and statistical errors,
that limit the  precision of its measurement.
Most importantly, the definition of $\sigma$ is problematic
 in the presence of string breaking effects, which would emerge in the presence of matter fields. 
 In this paper we 
 adopt an alternative strategy, 
 which could be adapted to more general gauge theories with fermionic matter field content.
 We will return to using $\sigma$ to set the scale for the topological
 susceptibility in Ref.~\cite{Bennett:2022gdz}, as it will help in the comparison with 
 measurements of $\chi$ in $SU(N_c)$ Yang-Mills theories.

The gradient flow  $B_\mu(x,\,t)$~\cite{Luscher:2010iy,Luscher:2013vga}
is defined unambiguously  in the continuum as on  the lattice,
with or without matter fields, and it can be determined precisely 
from simple averages of lattice observables. It is introduced as the 
solution to the  differential equation 
\beq
\frac{\mathrm{d} B_\mu(x,\,t)}{\mathrm{d}t} = 
D_\nu G_{\nu\mu}(x,\,t)\,,
\eeq
with boundary conditions $B_\mu(x,\,0)=A_\mu(x)$. The
independent variable $t$ is known as \emph{flow time},
while $D_\mu\equiv \partial_\mu + \left[B_\mu,\,\cdot\,\right]$, with
\beq
G_{\mu\nu}(t) = \left[D_\mu,\,D_\nu\right]\,.
\eeq

The defining properties of the gradient flow
make it suitable  as a smoothening procedure for UV fluctuations.
Since $\tfrac{d}{dt} S_\mathrm{YM}\leq 0$, a representative 
configuration $A_\mu(x)$ at $t=0$ is driven, along the flow,
towards a classical configuration.
In the perturbative regime, 
the flow equation
can be shown, at leading order in $g_0$,  
to generate a Gaussian smoothening operation with mean-square radius $\sqrt{8t}$. 
As a consequence, short-distance singularities in correlation functions of 
operators at $t > 0$ are eliminated.

The renormalised coupling $\alpha$ at scale $\mu=1/\sqrt{8t}$ is
\begin{equation}
\label{eq:flowE}
    \alpha(\mu) \equiv k_\alpha t^2 \langle E(t) \rangle 
    \equiv  k_\alpha \mathcal{E}(t)~,
\end{equation}
where $k_\alpha$ is a (perturbatively) calculable constant, and
\begin{equation}\label{eq:GF_coupling}
    E(t) \equiv \frac{1}{2} \mathrm{Tr} \,
    G_{\mu\nu}(t) G_{\mu\nu}(t)\,.
\end{equation}
The evolution of  $\alpha(1/\sqrt{8t})$ defines implicitly the scale $1/\sqrt{8t_0}$ of the system,
 by the requirement
\begin{equation}
\label{eq:scale_t0}
    \left. \mathcal{E}(t)\right|_{t=t_0} = \mathcal{E}_0\,,
\end{equation}
where $\mathcal{E}_0$ is a reference value, chosen for convenience.

Alternatively,  one 
defines  the observable~\cite{Borsanyi:2012zs} 
\begin{equation}\label{eq:flowW}
    {\cal W}(t) \equiv t \frac{d}{dt} \left\{ t^2 \langle \mathcal{E}(t) \rangle \right\}\,,
\end{equation}
and the scale $w_0$ is defined implicitly from
\begin{equation}
\label{eq:scale_w0}
   \left. {\cal W}(t)\right|_{t=w_0^2} = \mathcal{W}_0\,,
\end{equation}
where again $\mathcal{W}_0$ is a reference constant value. 
While $\mathcal{E}(t)$ is expected to be sensitive to the fluctuations of the gauge configurations on scales \emph{down to} $1/\sqrt{t_0}$, ${\cal W}(t)$ only depends on fluctuations \emph{around} $1/\sqrt{t_0}$.

We will compute the value of $t_0$ and $w_0$ in $Sp(N_c)$ theories for 
different values of $N_c$, with the implicit intention of exploring the 
$N_c\to\infty$ limit at constant 't Hooft coupling $\lambda\equiv 4\pi N_c\alpha$. 
From the  perturbative relation between $\mathcal{E}(t)$ and the 
gradient flow coupling~\cite{Luscher:2010iy}, we obtain 
the  leading-order expression
\beq
\mathcal{E}(t) = \frac{3
\lambda}{64\pi^2}~C_2(F)\,,
\label{Eq:ee}
\eeq
with $C_2(F)=\tfrac{N_c+1}{4}$, the quadratic Casimir operator of the fundamental representation
of $Sp(N_c)$.
 In order to compare different $Sp(N_c)$ theories, 
we  scale $\mathcal{E}_0$ and $\mathcal{W}_0$ according to the following relations:
\beq\label{eq:scale_scaling}
\mathcal{E}_0 (N_c) = c_e C_2(F),\quad \mathcal{W}_0 (N_c) = c_w C_2(F)\,,
\eeq
where $c_e$ and $c_w$ are constants. 
From Eq.~(\ref{Eq:ee}), leading-order perturbation theory gives
\beq
c_e = \frac{3\lambda}{64\pi^2}\,,
\eeq
showing that $c_e$ determines the fixed-$\lambda$ trajectory along which to take the $N_c\to\infty$
limit.\footnote{For unitary 
groups $SU(N_c)$, $C_2(F)=(N_c^2-1)/(2N_c)$. The choice $c_e=9/40$ would yield
$\mathcal{E}_0=0.3$ for $SU(3)$~\cite{Luscher:2010iy}.}
 Whether the scaling law in Eq.~(\ref{eq:scale_scaling}) holds outside of the domain of validity
of perturbation theory is a question we return to in Sec.~\ref{Sec:numerical}.

%%%%%%%%%%%%%%%%%
%%%%%%%%%%%%%%%%%
\subsection{The Wilson flow}
\label{Sec:Wilson}

The lattice incarnation of the gradient flow is based on the Wilson 
action ${S}_W$ in Eq.~(\ref{eq:Wilson_action}), and is known as the 
Wilson flow. $V_\mu(x,\,t)$ is 
defined by solving
\beq\label{eq:wilson_flow}
\frac{ \partial V_\mu(x,\,t) }
{\partial t} = -g_0^2 \left\{ 
\partial_{x,\,\mu} S^\mathrm{flow}\left[V_\mu\right]\right\}
V_\mu(x,\,t)\,,
\eeq
where $V_\mu(x,\,0)=U_\mu(x)$. 
The properties of the Wilson flow %in a discretised system 
are naturally inherited from the continuum formulation. 
Moreover, a numerical integration can be set up to obtain $V_\mu(x,\,t)$ from $U_\mu(x)$ explicitly, using for example a 
Runge-Kutta integration scheme, as detailed in Ref.~\cite{Luscher:2010iy}. 
Observables can then be constructed from $V_\mu(x,\,t)$.

To use the Wilson flow as a scale setting procedure 
 requires the computation of $\mathcal{E}(t)$ or $\mathcal{W}(t)$,
for which purpose two alternative lattice discretisations of $G_{\mu\nu}(t)$ can be used.
One is the elementary plaquette operator defined 
in Eq.~(\ref{eq:elementary_plaquette}), computed from $V_\mu(t)$. 
The other is the four-plaquette clover-leaf in Eq.~(\ref{Eq:clover}). 
By comparing numerically the values of $\mathcal{E}(t)$ 
and $\mathcal{W}(t)$, as well as the two different discretisations, 
we can assess the magnitude of discretisation errors in approaching the continuum limit.

%%%%%%%%%%%%%%%%%
%%%%%%%%%%%%%%%%%
\subsection{Topological susceptibility on the lattice}
\label{Sec:top}

As anticipated in Section~\ref{Sec:theory}, we compute the 
topological susceptibility on the lattice  from Wilson-flowed configurations.
At flow time $t$, the topological charge density can be obtained from
\begin{equation}
\label{eq:flowed_top_charge}
q_L(t,\,x) \equiv \frac{1}{32\pi^2} \epsilon^{\mu\nu\rho\sigma}
\mathrm{Tr} ~{\cal C}_{\mu\nu}(x,\,t) {\cal C}_{\rho\sigma}(x,\,t)\,,
\end{equation}
where ${\cal C}_{\mu\nu}(x,\,t)$ is the clover
operator computed from $V_\mu(x,\,t)$. 
The topological charge is
$Q_L(t)=\sum_x q_L(x,\,t)$. 

On the lattice, 
the values of the topological charge are quasi-integers, affecting  the
measurement  of $\chi_L$. Following Ref.~\cite{Bonati:2016tvi}, we  
reduce this systematical error by redefining $\tilde{Q}_L$ as follows:
\begin{equation}\label{eq:lat_top_charge_TOT}
    \tilde{Q}_L(t) \equiv {\rm round}\left(
    \tilde{\alpha} \sum_x q_L(x,\,t)\right)\,,
\eeq
where $\tilde{\alpha}$ is a numerical factor determined by minimising the $t$-dependent quantity
\begin{equation}
\label{eq:alpha}
\Delta(\tilde{\alpha})=\left\langle\left[ \tilde{\alpha} {Q}_L -
{\rm round} \left(\tilde{\alpha} {Q}_L\right)
\right]^2\right\rangle\,.
\eeq
We will provide an illustrative example of the choice of 
numerical factor $\tilde{\alpha}\sim {\cal O}(1)$
 when presenting our results.
The topological susceptibility at flow time $t$ is then
\begin{equation}\label{eq:flowed_topo_suscept}
\chi_L (t) a^4= \frac{1}{L^4}\left\langle \frac{}{}\tilde{Q}_L(t)^2 \frac{}{}\right\rangle~.
\end{equation}

%%%%%%%%%%%%%%%%%%%%%
%%%%%%%%%%%%%%%%%%%%%
\section{Numerical results}
\label{Sec:numerical}

\begin{table}
\caption{Ensembles used for  scale setting.
The first three columns show the bare parameters for each 
ensemble. $N_\textrm{tot}$ is the number of configurations,
and we applied $N_\textrm{sw}$ 
 lattice sweeps between two successive configurations.
The measurement of the integrated autocorrelation time 
 of the topological charge, $\tau_Q$, is discussed later in the paper. \label{tab:lattice_setting1}\\}
\centering
\begin{tabular}{|c|c|c|c|c|c|}
\hline
\hline
$~~~N_c~~~$ & $~~~L/a~~~$ & $~~~\beta~~~$ & $~~~N_\textrm{sw}~~~$ &~~~$N_\textrm{tot}$~~~ & $~~~\tau_Q~~~$ \\
\hline
$2$ & $20$ & $2.55$ & $50$ & $3999$ & $0.512(30)$ \\
$2$ & $24$ & $2.60$ & $100$ & $3999$ & $0.512(30)$ \\
$2$ & $32$ & $2.65$ & $100$ & $4003$ & $0.561(33)$ \\
$2$ & $32$ & $2.70$ & $100$ & $4003$ & $0.729(43)$ \\
\hline
$4$ & $20$ & $7.7$ & $50$ & $4000$ & $0.644(38)$ \\
$4$ & $20$ & $7.72$ & $50$ & $4000$ & $0.672(40)$ \\
$4$ & $20$ & $7.76$ & $50$ & $4000$ & $0.779(52)$ \\
$4$ & $20$ & $7.78$ & $40$ & $4002$ & $1.079(80)$ \\
$4$ & $20$ & $7.80$ & $80$ & $4021$ & $0.691(41)$ \\
$4$ & $20$ & $7.85$ & $70$ & $4002$ & $1.104(82)$ \\
$4$ & $24$ & $8.2$ & $3500$ & $3898$ & $0.550(33)$ \\
\hline
$6$ & $18$ & $15.75$ & $60$ & $4000$ & $0.848(57)$ \\
$6$ & $16$ & $15.9$ & $100$ & $4006$ & $0.959(64)$ \\
$6$ & $16$ & $16.1$ & $400$ & $4011$ & $1.170(87)$ \\
$6$ & $20$ & $16.3$ & $800$ & $4001$ & $1.47(12)$ \\
\hline
$8$ & $16$ & $26.5$ & $600$ & $3924$ & $0.617(37)$ \\
$8$ & $16$ & $26.7$ & $400$ & $4061$ & $1.27(10)$ \\
$8$ & $16$ & $27.0$ & $1200$ & $3887$ & $1.50(13)$ \\
$8$ & $16$ & $27.2$ & $3000$ & $4107$ & $1.245(99)$ \\
\hline
\end{tabular}
%
%\begin{tabular}{ |c |c |c| c| c| c|c|}
%\hline
%\hline
%$~~~N_c~~~$ & $~~~L/a~~~$ & $~~~\beta~~~$ & $~~~N_\textrm{tot}~~~$ &~~~$N_\textrm{sw}$~~~ & $~~~\tau_Q~~~$ \\
%\hline
%$2$ & $20$ & $2.55$ & $4000$ & $50$ & $0.510(30)$ \\
%$2$ & $24$ & $2.60$ & $4000$ & $100$ & $0.519(31)$ \\
%$2$ & $32$ & $2.65$ & $4000$ & $100$ & $0.563(33)$ \\
%$2$ & $32$ & $2.70$ & $4000$ & $100$ & $0.733(43)$ \\
%\hline                         
%$4$ & $20$ & $7.7$  & $4000$ &$   50$ & $0.639(38)$ \\
%$4$ & $20$ & $7.72$ & $3997$ &$   50$ & $0.655(39)$ \\
%$4$ & $20$ & $7.76$ & $4000$ &$   50$ & $0.778(52)$ \\
%$4$ & $20$ & $7.78$ & $4002$ &$   40$ & $1.082(80)$ \\
%$4$ & $20$ & $7.80$ & $4002$ &$   40$ & $1.35(11)$ \\
%$4$ & $20$ & $7.85$ & $4002$ &$   70$ & $1.113(83)$ \\
%$4$ & $24$ & $8.2$  & $3898$ &$ 3500$ & $0.550(33)$ \\
%\hline                         
%$6$ & $18$ & $15.75$& $3209$ &$   60$ & $0.891(67)$ \\
%$6$ & $16$ & $15.9$ & $4006$ &$  100$ & $0.955(64)$ \\
%$6$ & $16$ & $16.1$ & $4011$ &$  400$ & $1.168(87)$ \\
%$6$ & $20$ & $16.3$ & $4001$ &$  800$ & $1.47(12)$ \\
%\hline                         
%$8$ & $16$ & $26.5$ & $3925$ &$  600$ & $0.607(36)$ \\
%$8$ & $16$ & $26.7$ & $4061$ &$  400$ & $1.26(10)$ \\
%$8$ & $16$ & $27.0$ & $3888$ &$ 1200$ & $1.53(13)$ \\
%$8$ & $16$ & $27.2$ & $4107$ &$ 3000$ & $1.244(99)$ \\
%\hline
%\end{tabular}

\end{table}

We generated and stored ensembles of thermalised configurations 
 for the set of bare parameters listed in the three left-most columns of 
 Table~\ref{tab:lattice_setting1}. From a comparison with the results obtained for
$\sigma$ in Ref.~\cite{Bennett:2020qtj}, we know that $\sqrt{\sigma}L\geq 3$ 
for all such ensembles, and  neglect finite-volume effects.
In the following, we present and discuss our numerical results for the scale-setting procedure,
and for the topological susceptibility.
All data presented, as well as underlying raw data,
are available at Ref.~\cite{datapackage}, and the analysis code used to prepare
the main figures and tables are shared at Ref.~\cite{codepackage}.

%%%%%%%%%%%%%%%%%%%%%
\subsection{Setting the scale}

\begin{figure}[t]
\centering
\includegraphics{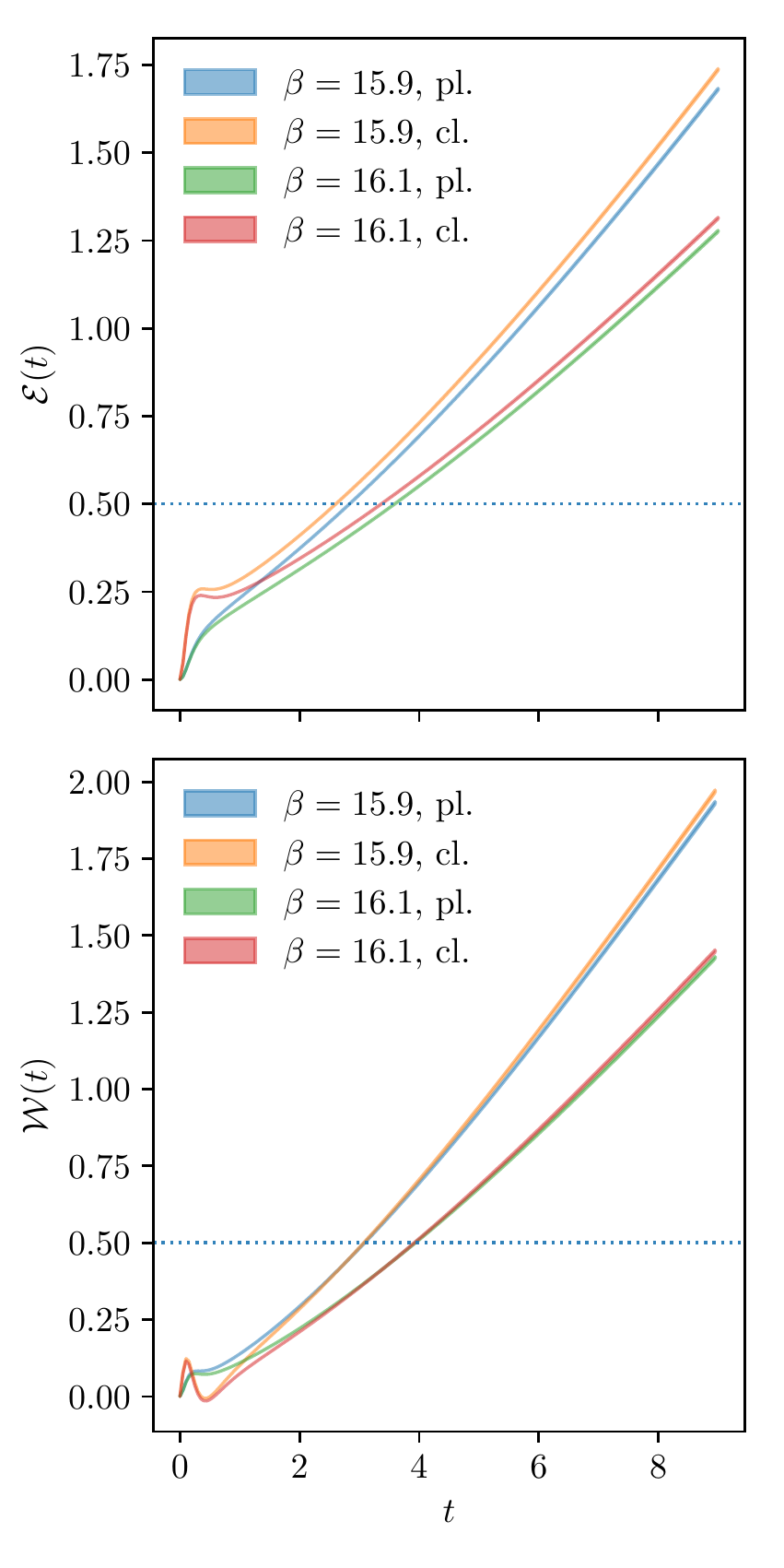}
\caption{The quantities $\mathcal{E}(t)$ (top panel) and $\mathcal{W}(t)$ (bottom panel), defined 
in Eqs.~(\ref{eq:flowE}) and~(\ref{eq:GF_coupling}),  and in Eq.~(\ref{eq:flowW}), respectively,
in the $N_c=6$ ensembles 
with $\beta=15.9$ and $\beta=16.3$,
as functions 
of the flow time $t$.  Computations adopt  the alternative choices
of discretisation provided by the  elementary plaquette (pl.) and the clover-leaf plaquette (cl.). 
The horizontal dashed line represents the choice ${\cal E}_0=0.5$, ${\cal W}_0=0.5$.\label{fig:nonscaled_flows}}
\end{figure}

\begin{figure}[t]
\centering
\includegraphics{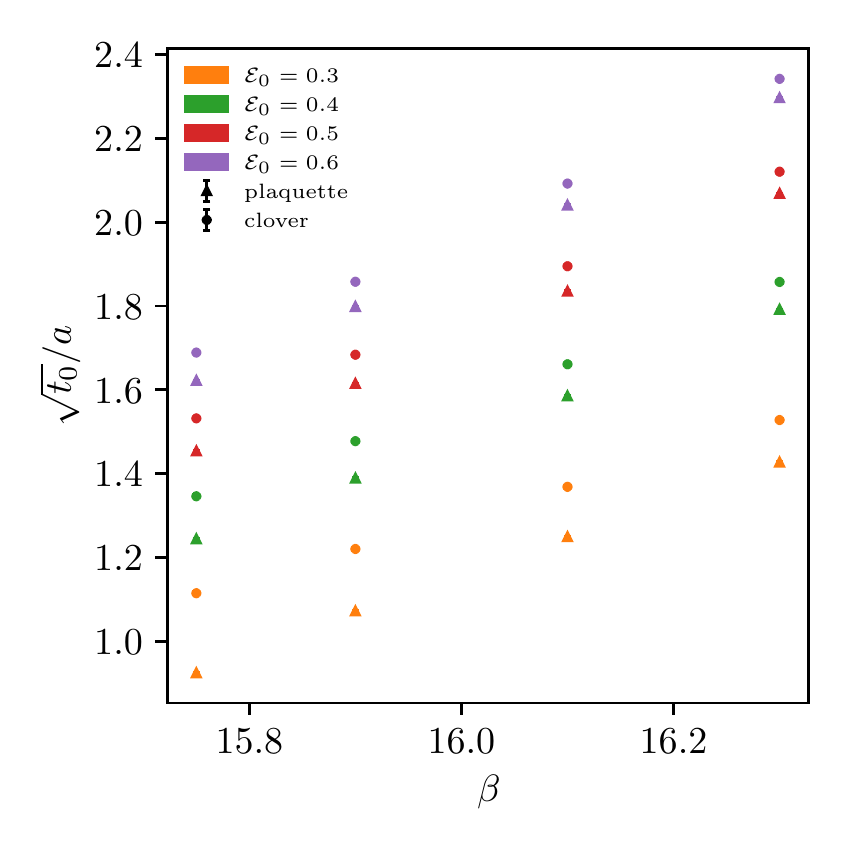}
\includegraphics{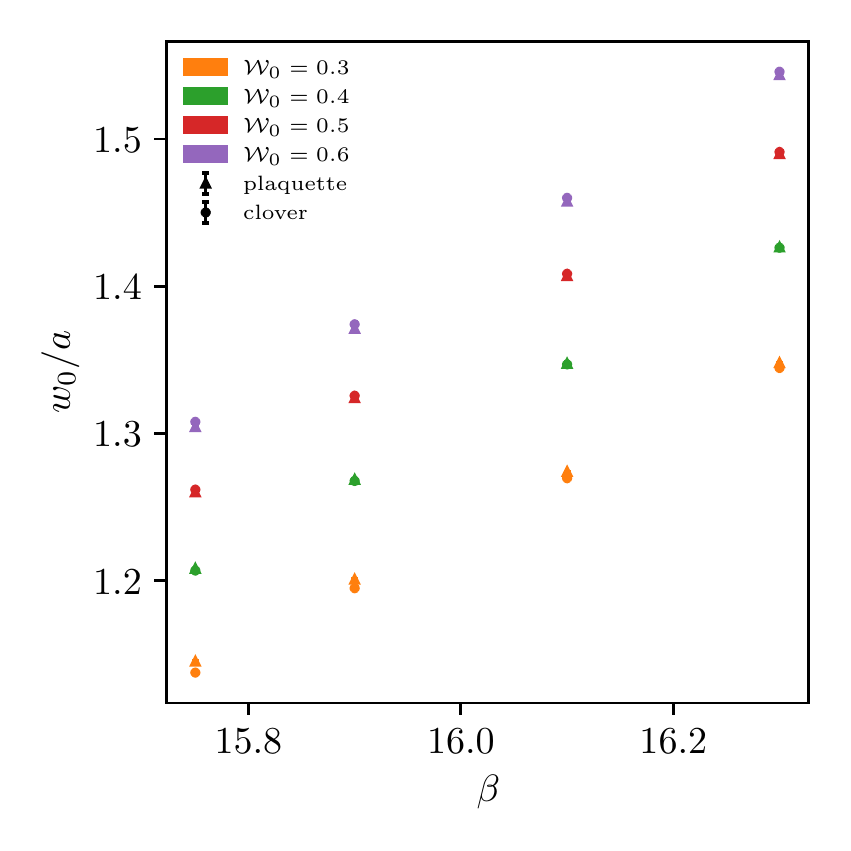}
\caption{The gradient flow scales $\sqrt{t_0}/a$ (top panel)
and  $w_0/a$  (bottom) in $Sp(6)$, 
for different choices of $\mathcal{E}_0$ or  ${\cal W}_0$,
as a function of $\beta$, and comparing plaquette and clover discretisation. 
In the case of $w_0/a$, the difference between discretisations is barely discernible over the statistical 
uncertainty.\label{fig:scale_6_t0_w0}
}
\end{figure}

Each configuration in the ensembles in  Table~\ref{tab:lattice_setting1}  sets the initial conditions
 for the numerical integration 
of the Wilson flow, which obeys Eq.~(\ref{eq:wilson_flow}). Following Ref.~\cite{Luscher:2010iy},
we use a third-order Runge-Kutta integrator (implemented by HiRep \cite{hirep-repo}) 
to evaluate $V_\mu(x,\,t)$ in
the range  $0<t<t_\text{max}$, with $t_\text{max}$ such that $\sqrt{8 t_\text{max}} \lesssim L/2$, to avoid large 
 finite size effects.

\begin{table}
\caption{The gradient flow scale $\sqrt{t_0}/a$ for different choices of  $\beta$
and $\mathcal{E}_0$,  for $N_c=6$. We report the results for both the plaquette 
and clover discretisations. \label{tab:scale_6_t0}\\}
\centering
\setlength{\tabcolsep}{10pt}
\begin{tabular}{|c|c|c|c|}
\hline
\hline
$\beta$ & $\mathcal{E}_0$ & $\sqrt{t_0}/a$ (pl.) & $\sqrt{t_0}/a$ (cl.) \\
\hline
$ 15.75 $&$ 0.3 $&$ 0.92442(25) $&$ 1.11491(20) $ \\
$ 15.9 $&$ 0.3 $&$ 1.07199(36) $&$ 1.22045(33) $ \\
$ 16.1 $&$ 0.3 $&$ 1.24921(48) $&$ 1.36866(47) $ \\
$ 16.3 $&$ 0.3 $&$ 1.42716(42) $&$ 1.52798(41) $ \\
$ 15.75 $&$ 0.4 $&$ 1.24407(31) $&$ 1.34614(29) $ \\
$ 15.9 $&$ 0.4 $&$ 1.38876(49) $&$ 1.47758(49) $ \\
$ 16.1 $&$ 0.4 $&$ 1.58507(71) $&$ 1.66099(69) $ \\
$ 16.3 $&$ 0.4 $&$ 1.79128(63) $&$ 1.85737(60) $ \\
$ 15.75 $&$ 0.5 $&$ 1.45355(39) $&$ 1.53196(37) $ \\
$ 15.9 $&$ 0.5 $&$ 1.61449(63) $&$ 1.68380(63) $ \\
$ 16.1 $&$ 0.5 $&$ 1.83487(92) $&$ 1.89487(89) $ \\
$ 16.3 $&$ 0.5 $&$ 2.06770(82) $&$ 2.12034(78) $ \\
$ 15.75 $&$ 0.6 $&$ 1.62169(46) $&$ 1.68890(44) $ \\
$ 15.9 $&$ 0.6 $&$ 1.79810(76) $&$ 1.85783(76) $ \\
$ 16.1 $&$ 0.6 $&$ 2.0401(11) $&$ 2.0920(11) $ \\
$ 16.3 $&$ 0.6 $&$ 2.2960(10) $&$ 2.34178(95) $ \\
\hline
\end{tabular}
\end{table}

\begin{table}
\caption{The gradient flow scale $w_0/a$ for different choices of  $\beta$ and $\mathcal{W}_0$,
 for $N_c=6$. We report the results for both the plaquette and clover discretisations. \label{tab:scale_6_w0}\\}
\centering
\setlength{\tabcolsep}{10pt}
\begin{tabular}{|c|c|c|c|}
\hline
\hline
$\beta$ & $\mathcal{W}_0$ & $w_0/a$ (pl.) & $w_0/a$ (cl.) \\
\hline
$ 15.75 $&$ 0.3 $&$ 1.14525(18) $&$ 1.13768(17) $ \\
$ 15.9 $&$ 0.3 $&$ 1.20091(29) $&$ 1.19500(29) $ \\
$ 16.1 $&$ 0.3 $&$ 1.27407(40) $&$ 1.26968(38) $ \\
$ 16.3 $&$ 0.3 $&$ 1.34793(34) $&$ 1.34454(32) $ \\
$ 15.75 $&$ 0.4 $&$ 1.20835(21) $&$ 1.20698(20) $ \\
$ 15.9 $&$ 0.4 $&$ 1.26876(33) $&$ 1.26784(33) $ \\
$ 16.1 $&$ 0.4 $&$ 1.34754(46) $&$ 1.34701(44) $ \\
$ 16.3 $&$ 0.4 $&$ 1.42652(39) $&$ 1.42624(37) $ \\
$ 15.75 $&$ 0.5 $&$ 1.26030(23) $&$ 1.26195(22) $ \\
$ 15.9 $&$ 0.5 $&$ 1.32419(37) $&$ 1.32572(37) $ \\
$ 16.1 $&$ 0.5 $&$ 1.40712(51) $&$ 1.40852(48) $ \\
$ 16.3 $&$ 0.5 $&$ 1.48995(43) $&$ 1.49123(41) $ \\
$ 15.75 $&$ 0.6 $&$ 1.30458(25) $&$ 1.30797(24) $ \\
$ 15.9 $&$ 0.6 $&$ 1.37126(40) $&$ 1.37421(40) $ \\
$ 16.1 $&$ 0.6 $&$ 1.45758(55) $&$ 1.46009(53) $ \\
$ 16.3 $&$ 0.6 $&$ 1.54353(47) $&$ 1.54572(45) $ \\
\hline
\end{tabular}
\end{table}

The quantity ${\cal E}(t)$ is obtained from the definitions in 
Eqs.~(\ref{eq:flowE}) and~(\ref{eq:GF_coupling}), 
by computing $G_{\mu\nu}$ from $B_\mu$.
We use two alternative discretised expressions for ${\cal E}(t)$,   provided by 
the plaquette (pl.), ${\cal P}_{\mu\nu}$, or the clover-leaf (cl.), 
${\cal C}_{\mu\nu}$. We then compute ${\cal W}(t)$ 
according to Eq.~(\ref{eq:flowW}).
The resampled results for  ${\cal E}(t)$ and ${\cal W}(t)$, as functions of $t$,
are displayed in the two panels of
Fig.~\ref{fig:nonscaled_flows}, in the case $N_c=6$, with $\beta=15.9$ and $\beta=16.3$ and for the two different 
discretizations (pl. and cl.).  For each value of $t$, the vertical
thickness of the curves represents the error of 
${\cal E}(t)$ and ${\cal W }(t)$, computed by bootstrapping. 
The picture is qualitatively the 
same for other values of the
bare parameters and is similar to the $SU(N_c)$ case. 

\begin{figure}[t]
\centering
\includegraphics{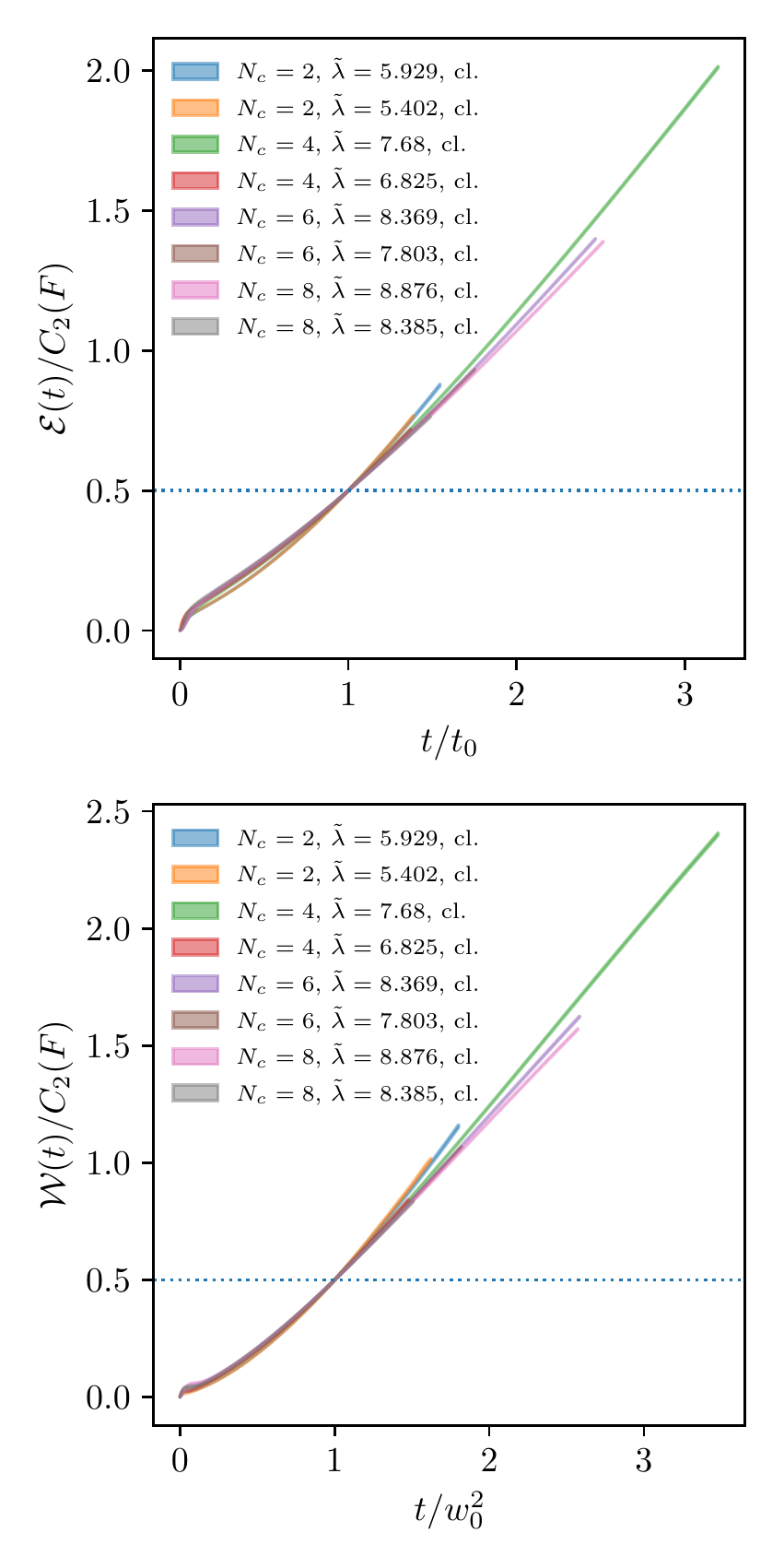}
\caption{
The quantities $\mathcal{E}(t)/C_2(F)$ (top panel) and
$\mathcal{W}(t)/C_2(F)$ (bottom) computed with the clover-leaf plaquette discretisation of the flow equation 
on the available ensembles corresponding to 
the finest and coarsest available lattices, for each $N_c$, with $C_2(F)=(N_c+1)/4$, displayed as a function
of the rescaled flow times $t/t_0$ and $t/w_0^2$.  The figure adopts the choice  $c_e=c_w=0.5$
(horizontal dashed line).
\label{fig:scaled_flows_0.5}}
\end{figure}

\begin{figure}[t]
\centering
\includegraphics{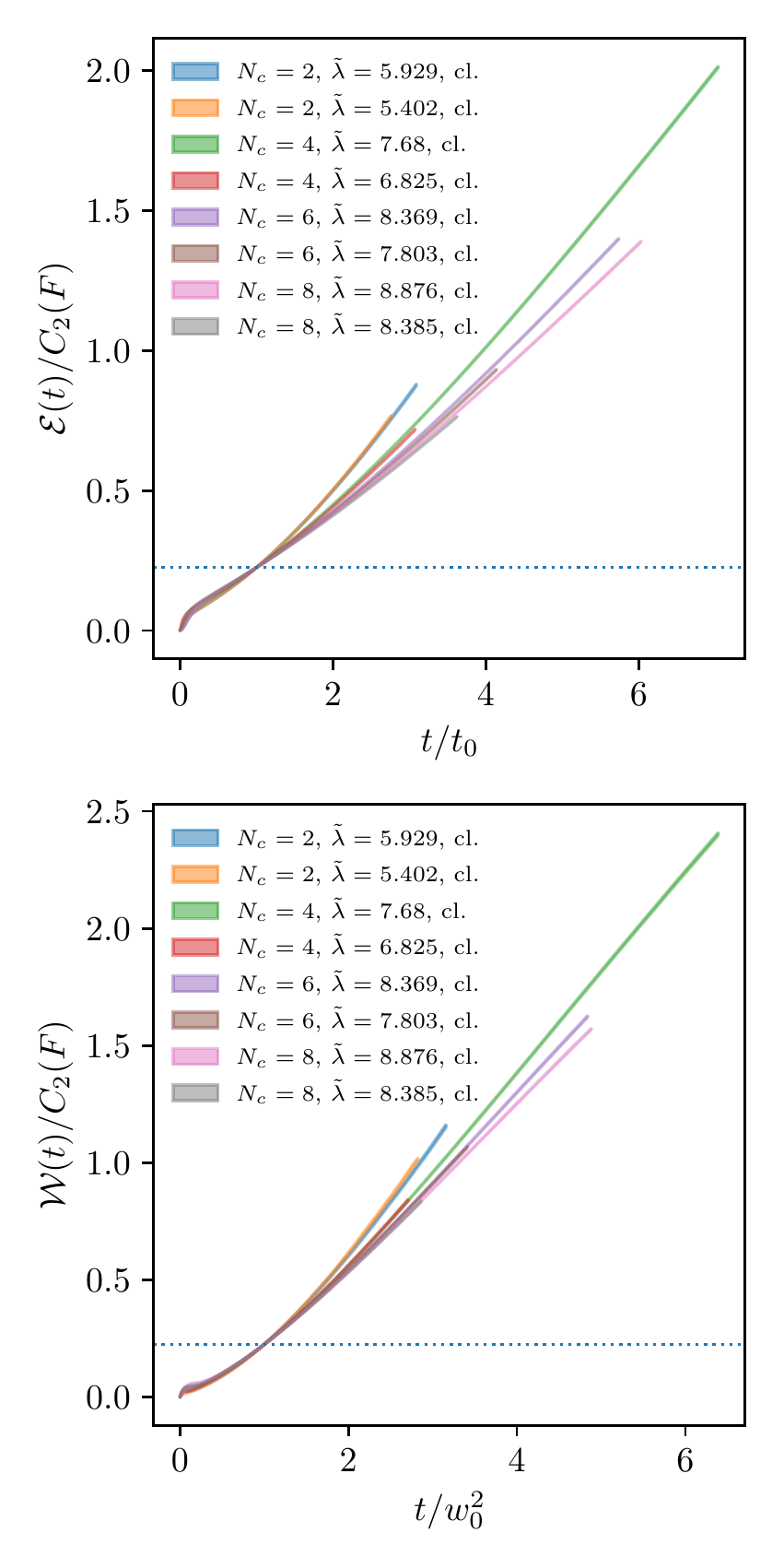}
\caption{
The quantities $\mathcal{E}(t)/C_2(F)$ (top panel) and
$\mathcal{W}(t)/C_2(F)$ (bottom) computed with the clover-leaf plaquette discretisation of the flow equation 
on the available ensembles corresponding to 
the finest and coarsest lattices, for each $N_c$, with $C_2(F)=(N_c+1)/4$, displayed as a function
of the rescaled flow times $t/t_0$ and $t/w_0^2$.  The figure adopts the choice  $c_e=c_w=0.225$
(horizontal dashed line). 
\label{fig:scaled_flows_0.225}
}
\end{figure}

From ${\cal E}(t)$ and ${\cal W}(t)$ we can
extract the scales $t_0$ and $w_0$, according to the definitions in
Eqs.~(\ref{eq:scale_t0}) and~(\ref{eq:scale_w0}), once we make a choice for
the reference values ${\cal E}_0$ and ${\cal W}_0$.
For  illustrative purposes, the choices $\mathcal{E}_0=0.5$ and 
$\mathcal{W}_0=0.5$ are represented as horizontal dashed lines in the top
and bottom panel of Fig.~\ref{fig:nonscaled_flows}, though we do not
use this choice in the analysis.
The comparison 
between  ${\cal E}(t)$ and 
${\cal W}(t)$ 
provides a first assessment of the magnitude of discretisation effects
in the calculation. For each ensemble, the 
difference between the curves corresponding to the
plaquette and clover discretisations tends to a constant
at large $t$. This difference is the smallest in the ensemble
with the largest value of $\beta$---the closest to the continuum limit---and
 is
smaller for ${\cal W}(t)$ than ${\cal E}(t)$, as anticipated in Section~\ref{Sec:scale_setting}.

A more refined assessment of discretisation effects 
can be obtained by studying the value of the scales $\sqrt{t_0}/a$ and $w_0/a$, obtained 
for a range of choices of ${\cal E}_0$ and ${\cal W}_0$, for each available ensemble, 
which we report
in Tables~\ref{tab:scale_6_t0} and~\ref{tab:scale_6_w0}, and  display 
 in Fig.~\ref{fig:scale_6_t0_w0} for $Sp(6)$.
The difference 
between the plaquette and clover discretizsations becomes smaller as
$\beta$ is increased. This difference has generally a smaller
magnitude for the $w_0$ scale than for  $\sqrt{t_0}$ scale. 
A similar picture emerges for  $N_c=2$, $N_c=4$, and $N_c=8$. 
as reported in  Appendix~\ref{Appendix}. 
In view of these considerations, we adopt the clover discretisation in
the following.

\begin{figure}[t]
\centering
\includegraphics[width=.45\textwidth]{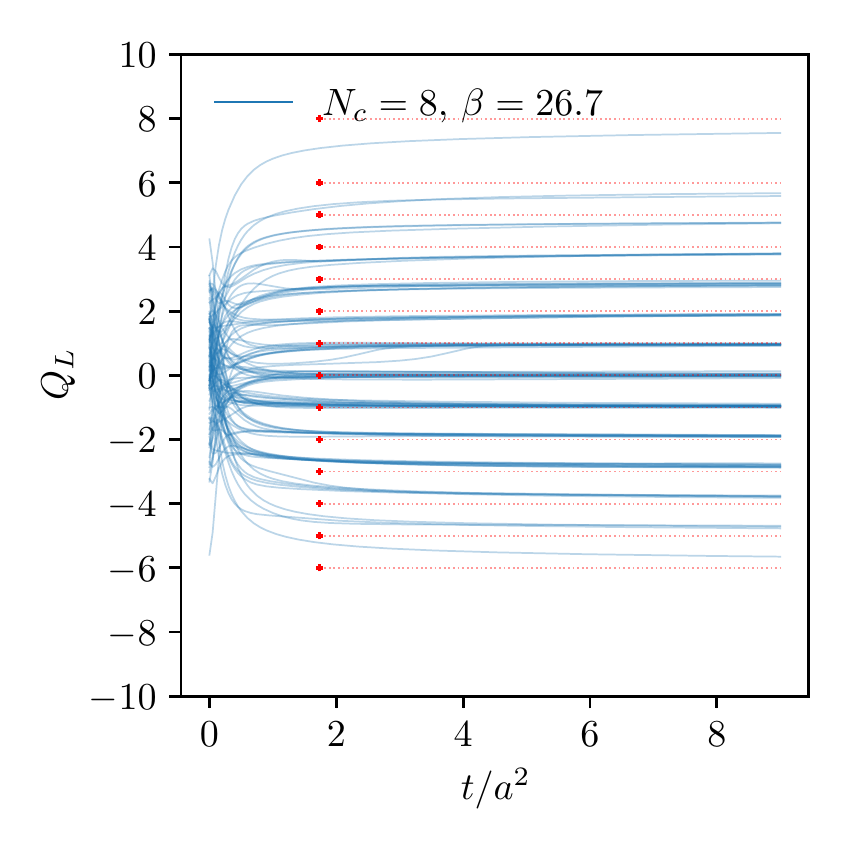}
\includegraphics[width=.45\textwidth]{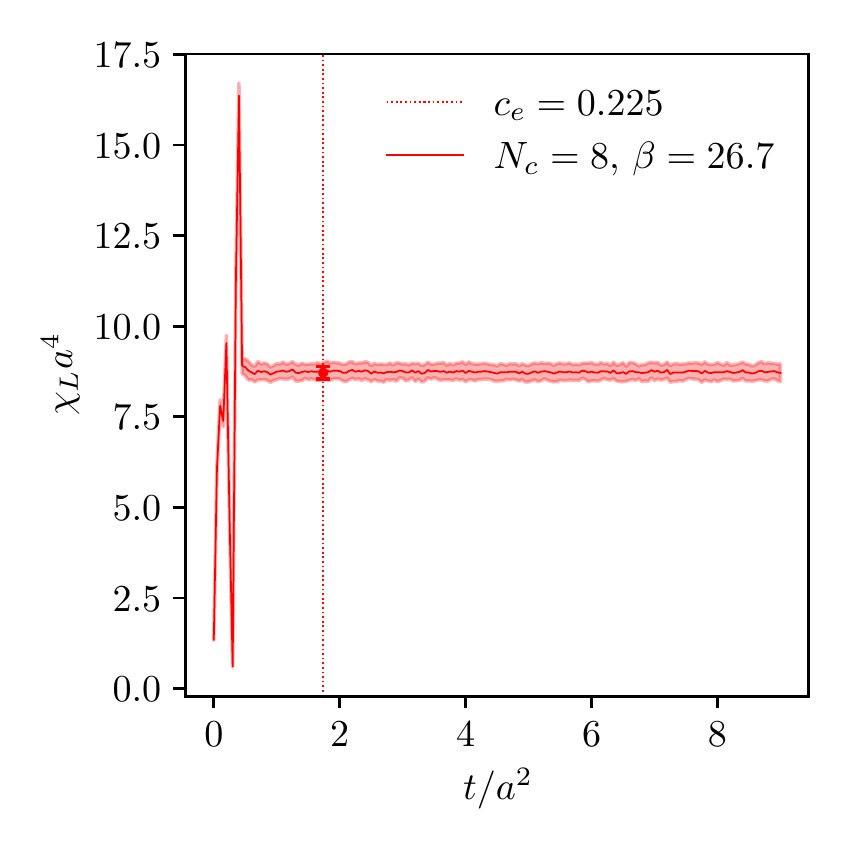}
\caption{In the top panel, he topological charge $Q_L$ as a function of the flow time $t/a^2$, for
the $100$ first configurations of the ensemble with $N_c=8$, $L/a=16$,  and
$\beta=26.7$. The values of $\tilde{Q}_L$ are also reported as red bullets
at $t=t_0$, where $t_0$ is obtained from the choice $c_e=0.225$.
In the bottom panel, the topological susceptibility $\chi_L(t)a^4$
as a function of the flow time $t$, with its $1$-$\sigma$ error band,
for $N_c=8$, at $\beta=26.7$. The vertical dashed
represents $t=t_0$, where $t_0$ is obtained from $c_e=c_w=0.225$.
The value of $\chi_L(t=t_0)a^4$ obtained at $t=t_0$ is depicted as a red bullet and reported in
Table~\ref{tab:top_suscept_0.225_t}.
\label{fig:TC_vs_t}}
\end{figure}

In Section~\ref{Sec:scale_setting}, we observed that the functions ${\cal E}(t)$ and 
${\cal W}(t)$ obey perturbative relations that suggest  the scaling bahaviour in Eq.~(\ref{eq:scale_scaling}).
The numerical data we have collected allows to test the validity of these
scaling relations outside of the domain of perturbation theory.
We choose values of $\mathcal{E}_0$ and $\mathcal{W}_0$ for
each value of $N_c$ according to Eq.~(\ref{eq:scale_scaling}), with fixed $c_e$ and $c_w$. 
The corresponding values of $t_0$ and $w_0$ are then
used to scale also $t$.

The result of these operations, with the  clover-leaf discretisation, 
and at each fixed value of $N_c$,  for the ensembles
with the smallest and the greatest  values of $\beta$,
 are 
displayed in Fig.~\ref{fig:scaled_flows_0.5}
 for
the choice $c_e=c_w=0.5$. The plots exhibit the same qualitative features
for both ${\cal E}(t)/C_2(F)$ and ${\cal W}(t)/C_2(F)$ as functions of $t/t_0$ 
and $t/w_0^2$, respectively. 
We repeat the same process also for the choice $c_e=c_w=0.225$ (which would yield ${\cal E}_0=0.3$ for $SU(3)$),
and display the results in Fig.~\ref{fig:scaled_flows_0.225}.
We labelled the curves by the conveniently defined discretised coupling
\beqs
\tilde{\lambda}&\equiv&\frac{d_G}{\beta\left\langle\frac{ \Re \Tr \mathcal{P}_{\mu\nu} }{2N}\right\rangle}\,,
\eeqs
where $d_G$ is the dimension of the group.

By construction,  the rescaled flows
corresponding to different values of $N_c$ coincide
when $t/t_0=1$ (or $t/w_0^2=1$). 
What is interesting is that the curves agree (within uncertainties)
 over a sizeable range of $t$ around these points. This might indicate that 
  the validity of the perturbative
scaling in Eq.~(\ref{eq:scale_scaling}) holds also outside of the naive range of the
perturbative regime.
Small deviations from perfect scaling are nevertheless visible
and might be ascribed to a combination of finite-$a$ and 
finite-$N_c$ effects.

\begin{figure}[t]
\centering
\includegraphics{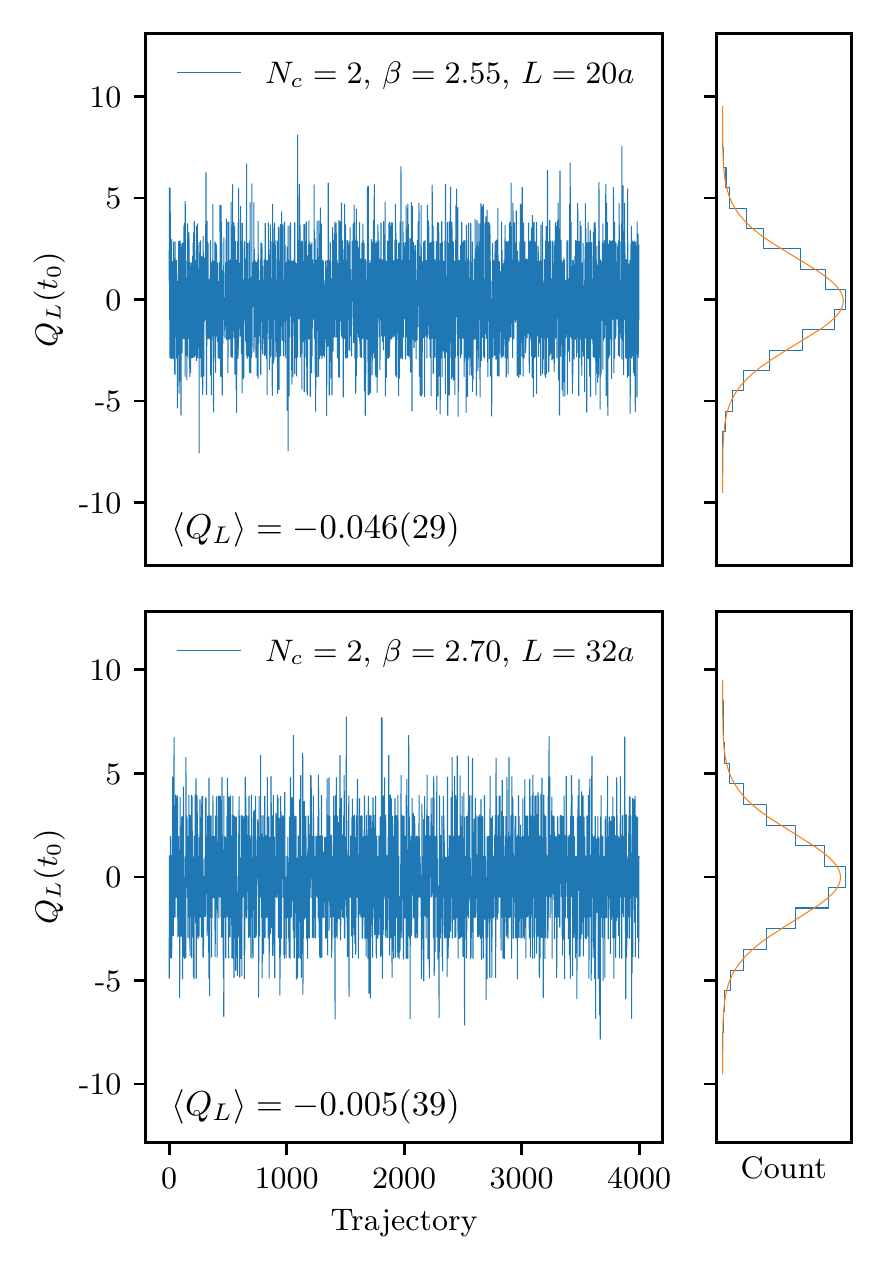}
\caption{The topological charge $Q_L$  as a function of simulation 
time (trajectory) for the ensembles corresponding to the coarsest (top) and finest (bottom) lattice
with $N_c=2$. 
The value of $Q_L$ is computed at $t=t_0$, where the value of $t_0$ is obtained from $c_e=0.225$. 
The average value of the topological charge along the trajectory
is reported in the bottom left-hand side of plot. 
The side panel contains the cumulative histogram 
of the values of $Q_L(t_0)$.
The orange curve is a gaussian fit to the cumulative
histogram.
\label{fig:TC_hist_2}}
\end{figure}

%%%%%%%%%%%%%%%%%%%%%%%%%%%%%%%%%%%%%
\subsection{The topological charge}

\begin{figure}[t]
\centering
\includegraphics{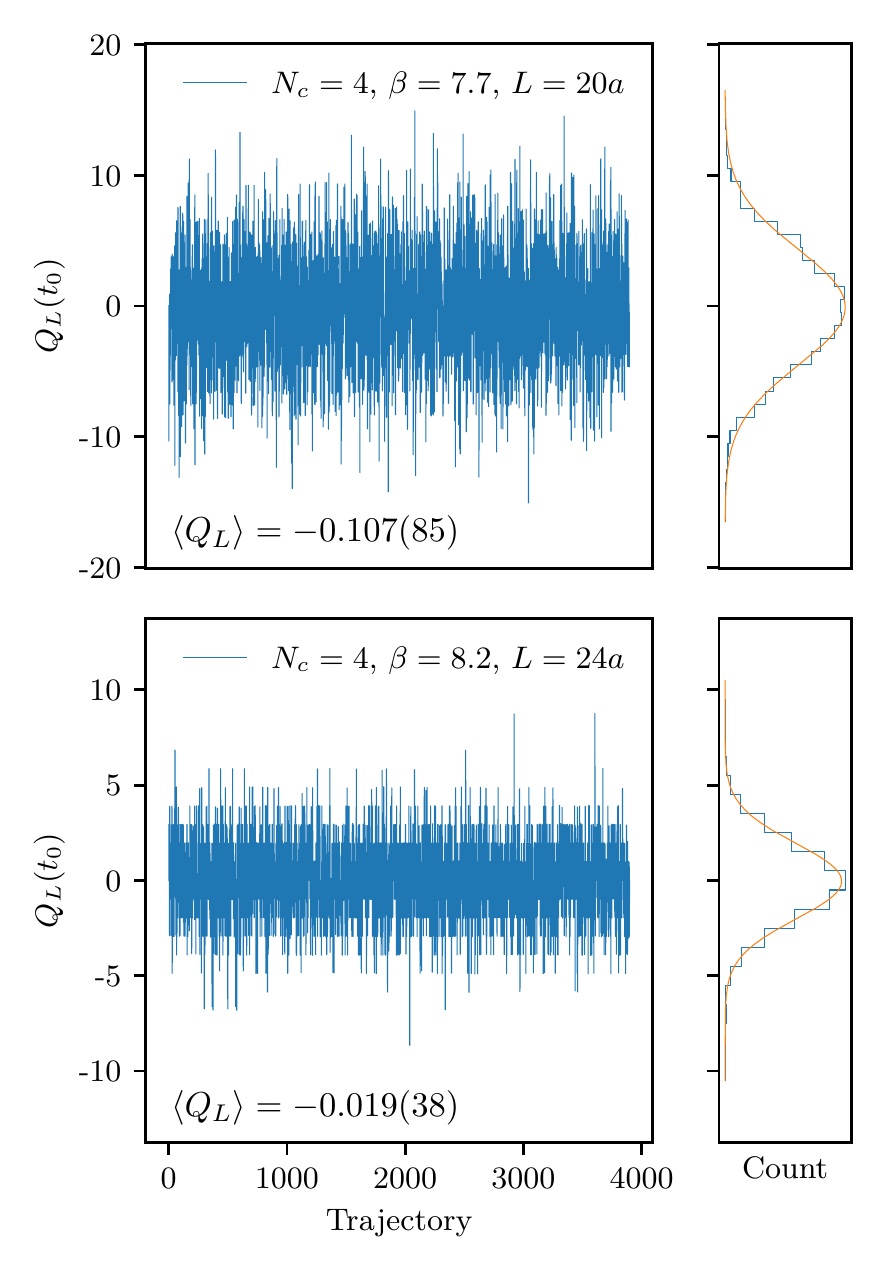}
\caption{The topological charge $Q_L$  as a function of simulation 
time (trajectory) for the ensembles corresponding to the coarsest (top) and finest (bottom) lattice
with $N_c=4$. 
The value of $Q_L$ is computed at $t=t_0$, where the value of $t_0$ is obtained from $c_e=0.225$. 
The average value of the topological charge along the trajectory
is reported in the bottom left-hand side of plot. 
The side panel contains the cumulative histogram 
of the values of $Q_L(t_0)$.
The orange curve is a gaussian fit to the cumulative
histogram.
\label{fig:TC_hist_4}}
\end{figure}

\begin{figure}[t]
\centering
\includegraphics{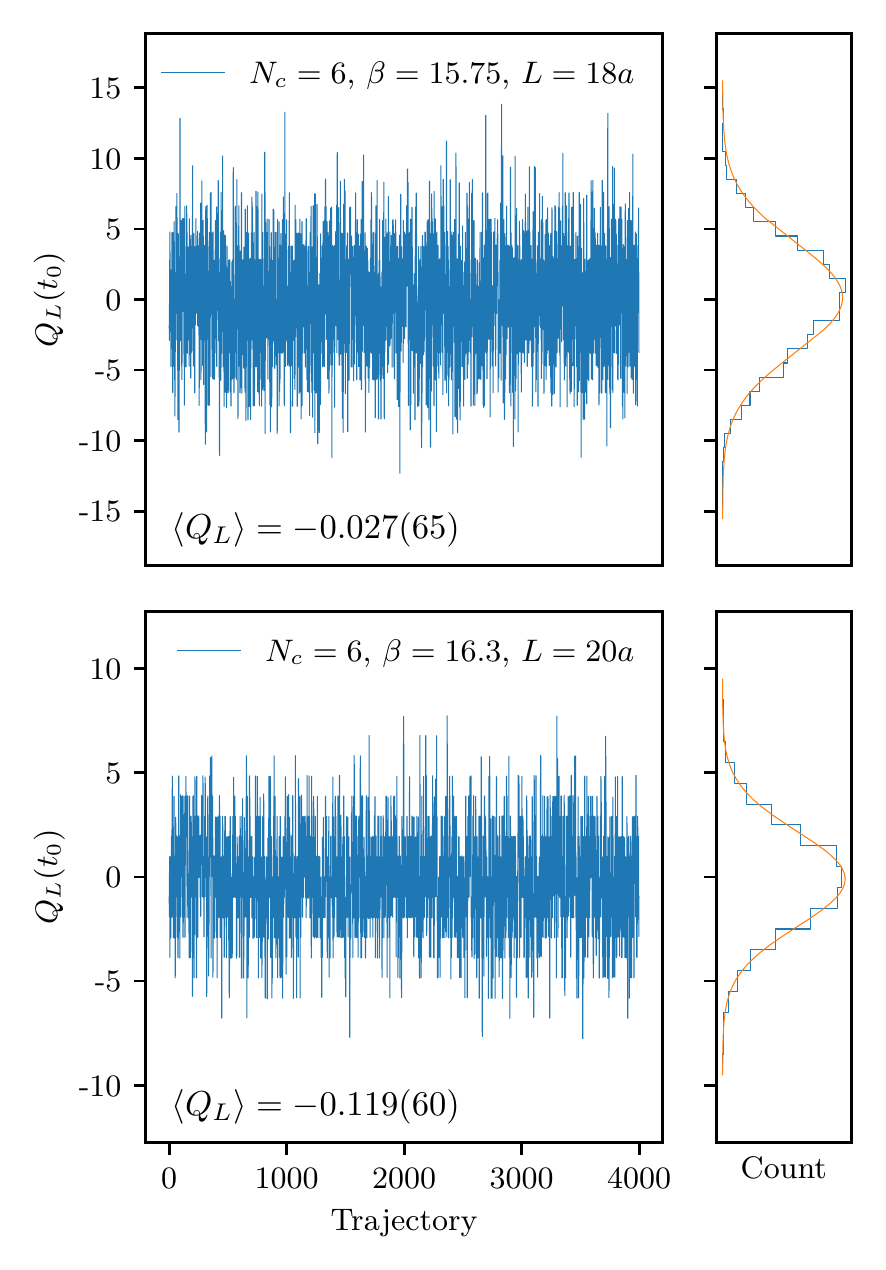}
\caption{The topological charge $Q_L$  as a function of simulation 
time (trajectory) for the ensembles corresponding to the coarsest (top) and finest (bottom) lattice
with $N_c=6$. 
The value of $Q_L$ is computed at $t=t_0$, where the value of $t_0$ is obtained from $c_e=0.225$. 
The average value of the topological charge along the trajectory
is reported in the bottom left-hand side of plot. 
The side panel contains the cumulative histogram 
of the valuoThe orange curve is a gaussian fit to the cumulative 
histogram. es of $Q_L(t_0)$.
\label{fig:TC_hist_6}}
\end{figure}

\begin{figure}[t]
\centering
\includegraphics{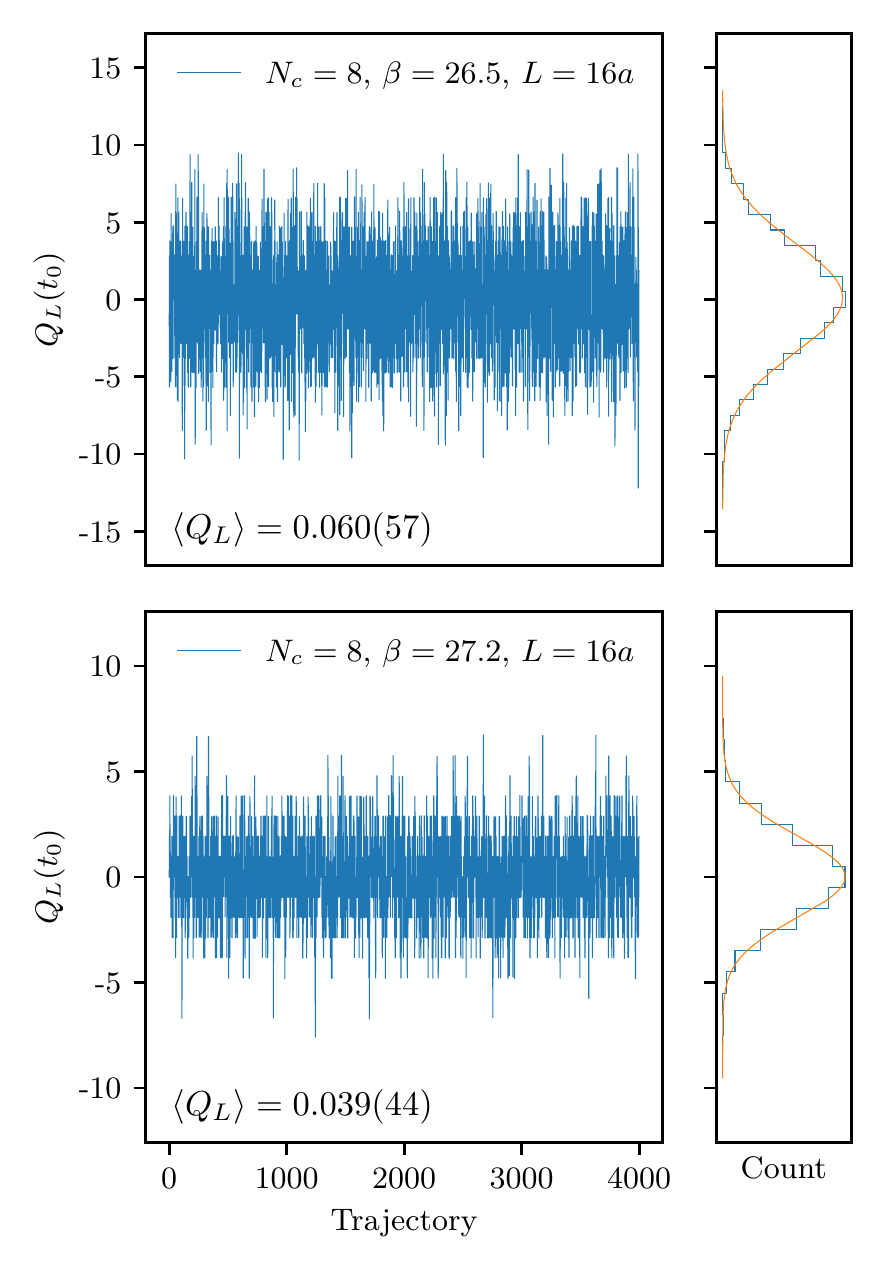}
\caption{The topological charge $Q_L$  as a function of simulation 
time (trajectory) for the ensembles corresponding to the coarsest (top) and finest (bottom) lattice
with $N_c=8$. 
The value of $Q_L$ is computed at $t=t_0$, where the value of $t_0$ is obtained from $c_e=0.225$. 
The average value of the topological charge along the trajectory
is reported in the bottom left-hand side of plot. 
The side panel contains the cumulative histogram 
of the values of $Q_L(t_0)$.
The orange curve is a gaussian fit to the cumulative
histogram.
\label{fig:TC_hist_8}}
\end{figure}

\begin{figure}[t]
\centering
\includegraphics{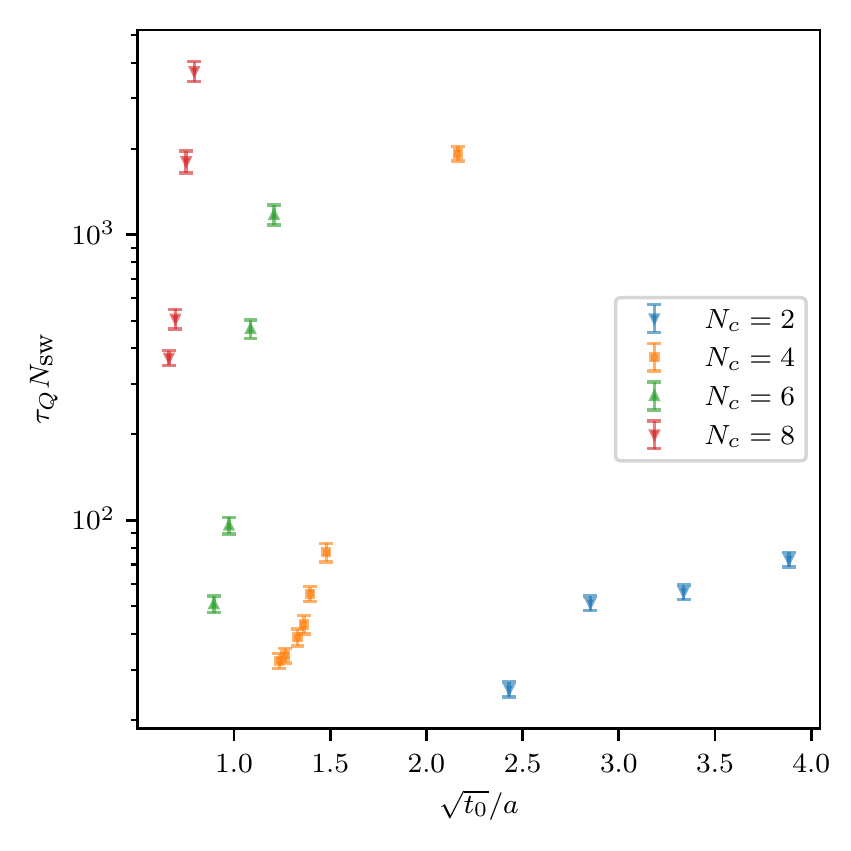}
\caption{The integrated autocorrelation time $\tau_Q$ multiplied by the number of sweeps $N_{\rm sw}$,
as a function of the scale $\sqrt{  t_0}/a$, for $c_e=0.225$ for $N_c=2$, $4$, $6$, $8$.\label{fig:tau_Q}}
\end{figure}

The quantity $Q_L(t)$ is computed from Wilson-flowed configurations
at flow time $t$, using Eq.~(\ref{eq:flowed_top_charge}), as implemented by HiRep \cite{hirep-repo}. Fig.~\ref{fig:TC_vs_t} 
is compiled with the first 100 configurations of the ensemble with $N_c=8$ and $\beta=26.7$.
Integer values for $Q_L(t)$ are only obtained for
$a\to 0$. At non-zero $a$, the quantity $Q_L(t)$ tends,
at large $t$, towards quasi-integer values.
The integer-valued topological charge
$\tilde{Q}_L$ is instead obtained for a finite value of $t$,
according to 
Eq.~(\ref{eq:lat_top_charge_TOT}). The values of $\tilde{Q}_L$ are displayed in 
Fig.~\ref{fig:TC_vs_t} as red bullets, and for the value of $t=t_0$ identified with the choice $c_e=0.225$. 
Effectively this definition of $\tilde{Q}_L$
 optimises and expedites the convergence towards the physical, discrete values
of the topological charge.
Similar conclusions hold for
 other values of the bare parameters, though we do not report further details here.
In the following, we will compute $\tilde{Q}_L$ for every 
configuration of every ensemble at both $c_e=0.225$ and $c_e=0.5$.

Simulation histories of $Q_L$ for the ensembles with the finest and coarsest lattice are displayed in 
Figs.~\ref{fig:TC_hist_2}, \ref{fig:TC_hist_4}, \ref{fig:TC_hist_6}, and~\ref{fig:TC_hist_8},
 for $N_c=2,\,4,\,6,\,8,$ respectively. 
The frequency histogram for the values of $Q_L$ is also reported, in each case,
and is consistent with a Gaussian,  symmetric distribution around
$Q_L=0$.

The magnitude
of autocorrelations can be evaluated using  the Madras-Sokal windowing algorithm~\cite{Madras:1988ei}
on the simulation history of either $Q_L$ or $\tilde{Q}_L$. 
For our ensembles, 
we found no significant difference to arise between the two,
and no visible dependence on the choice  of $c_e$.
We thus compute $\tau_Q$ at $t=t_0$, for $c_e=0.225$. 
The behaviour of the logarithm of $\tau_Q N_{sw}$ as a function of $\beta$ is displayed in 
Fig.~\ref{fig:tau_Q} and reported in Table~\ref{tab:lattice_setting1}. 
On the basis of known results obtained for
 $SU(N_c)$ gauge groups, $\ln \tau_Q N_{sw}$ is expected  to be linearly diverging
as $\beta\to\infty$~\cite{DelDebbio:2002xa}, and indeed this is consistent with  what we find.

\begin{table}
\centering
\caption{Topological susceptibilities $\chi_L t_0^2$  and $\chi_L w_0^4$, with $c_e=c_w=0.225$, and the string tensions $\sigma t_0$ and $\sigma w_0^2$, for all  
available ensembles. These results are displayed as a function of  $a^2/{t_0}$ and $a^2/w_0^2$  in Figure~\ref{fig:contlim}.\label{tab:top_suscept_0.225_t}\\}
\begin{tabular}{|cccccc|}
\hline
\hline
$N_c$ & $\beta$ & $\sigma t_0$ & $\chi_L t_0^2 \cdot 10^4$ &$\sigma w_0^2$ & $\chi_L w_0^4 \cdot 10^4$ \\
\hline
$ 2 $ & $ 2.55 $ & $ 0.11440(93) $ & $ 6.69(16) $ & $ 0.11164(92) $ & $ 6.40(15) $ \\
$ 2 $ & $ 2.6 $ & $ 0.1120(11) $ & $ 6.39(12) $ & $ 0.1090(10) $ & $ 6.07(14) $ \\
$ 2 $ & $ 2.65 $ & $ 0.1120(20) $ & $ 6.55(14) $ & $ 0.1092(19) $ & $ 6.21(14) $ \\
$ 2 $ & $ 2.7 $ & $ 0.1097(18) $ & $ 6.17(14) $ & $ 0.1070(18) $ & $ 5.87(17) $ \\
\hline
$ 4 $ & $ 7.7 $ & $ 0.1029(24) $ & $ 5.19(12) $ & $ 0.1127(26) $ & $ 6.47(15) $ \\
$ 4 $ & $ 7.72 $ & $ 0.1035(28) $ & $ 5.52(12) $ & $ 0.1138(31) $ & $ 6.80(16) $ \\
$ 4 $ & $ 7.76 $ & $ 0.0987(29) $ & $ 5.50(11) $ & $ 0.1087(32) $ & $ 6.71(16) $ \\
$ 4 $ & $ 7.78 $ & $ 0.1028(29) $ & $ 5.29(15) $ & $ 0.1136(32) $ & $ 6.44(15) $ \\
$ 4 $ & $ 7.8 $ & $ 0.0988(23) $ & $ 5.06(12) $ & $ 0.1093(26) $ & $ 6.16(15) $ \\
$ 4 $ & $ 7.85 $ & $ 0.1013(14) $ & $ 5.18(13) $ & $ 0.1125(16) $ & $ 6.40(16) $ \\
$ 4 $ & $ 8.2 $ & $ 0.1036(16) $ & $ 4.79(12) $ & $ 0.1171(18) $ & $ 6.11(15) $ \\
\hline
$ 6 $ & $ 15.75 $ & $ 0.0972(12) $ & $ 4.61(12) $ & $ 0.1146(15) $ & $ 6.36(15) $ \\
$ 6 $ & $ 15.9 $ & $ 0.0988(13) $ & $ 4.244(93) $ & $ 0.1177(15) $ & $ 5.98(13) $ \\
$ 6 $ & $ 16.1 $ & $ 0.0947(16) $ & $ 4.19(10) $ & $ 0.1138(20) $ & $ 6.06(15) $ \\
$ 6 $ & $ 16.3 $ & $ 0.0955(33) $ & $ 3.85(10) $ & $ 0.1153(40) $ & $ 5.61(15) $ \\
\hline
$ 8 $ & $ 26.5 $ & $ 0.0953(13) $ & $ 4.146(78) $ & $ 0.1170(16) $ & $ 6.27(11) $ \\
$ 8 $ & $ 26.7 $ & $ 0.0954(26) $ & $ 4.03(11) $ & $ 0.1181(32) $ & $ 6.16(17) $ \\
$ 8 $ & $ 27.0 $ & $ 0.0942(14) $ & $ 3.99(10) $ & $ 0.1177(17) $ & $ 6.23(18) $ \\
$ 8 $ & $ 27.2 $ & $ 0.0905(13) $ & $ 3.606(90) $ & $ 0.1138(17) $ & $ 5.72(18) $ \\

\hline
\end{tabular}

\end{table}

\begin{table}
\centering
\caption{Topological susceptibilities $\chi_L t_0^2$  and $\chi_L w_0^4$, with $c_e=c_w=0.5$, and the string tensions $\sigma t_0$ and $\sigma w_0^2$, for all  
available ensembles. These results are displayed as a function of  $a^2/{t_0}$ and $a^2/w_0^2$  in Figure~\ref{fig:contlim}.
\label{tab:top_suscept_0.5_t}\\}
\begin{tabular}{|cccccc|}
\hline
\hline
$N_c$ & $\beta$ & $\sigma t_0$ & $\chi_L t_0^2 \cdot 10^3$ &$\sigma w_0^2$ & $\chi_L w_0^4 \cdot 10^3$ \\
\hline
$ 2 $ & $ 2.55 $ & $ 0.2289(19) $ & $ 2.547(58) $ & $ 0.1956(16) $ & $ 1.881(47) $ \\
$ 2 $ & $ 2.6 $ & $ 0.2235(21) $ & $ 2.448(49) $ & $ 0.1907(18) $ & $ 1.789(35) $ \\
$ 2 $ & $ 2.65 $ & $ 0.2233(39) $ & $ 2.527(60) $ & $ 0.1902(33) $ & $ 1.840(40) $ \\
$ 2 $ & $ 2.7 $ & $ 0.2186(37) $ & $ 2.388(63) $ & $ 0.1861(31) $ & $ 1.736(45) $ \\
\hline
$ 4 $ & $ 7.7 $ & $ 0.2265(52) $ & $ 2.510(60) $ & $ 0.2070(47) $ & $ 2.100(51) $ \\
$ 4 $ & $ 7.72 $ & $ 0.2281(62) $ & $ 2.669(76) $ & $ 0.2087(56) $ & $ 2.246(56) $ \\
$ 4 $ & $ 7.76 $ & $ 0.2178(63) $ & $ 2.647(59) $ & $ 0.1992(58) $ & $ 2.225(46) $ \\
$ 4 $ & $ 7.78 $ & $ 0.2271(64) $ & $ 2.556(61) $ & $ 0.2080(59) $ & $ 2.137(57) $ \\
$ 4 $ & $ 7.8 $ & $ 0.2186(51) $ & $ 2.430(53) $ & $ 0.2001(47) $ & $ 2.040(48) $ \\
$ 4 $ & $ 7.85 $ & $ 0.2246(31) $ & $ 2.507(67) $ & $ 0.2060(29) $ & $ 2.113(53) $ \\
$ 4 $ & $ 8.2 $ & $ 0.2323(36) $ & $ 2.399(62) $ & $ 0.2146(33) $ & $ 2.055(47) $ \\
\hline
$ 6 $ & $ 15.75 $ & $ 0.2261(29) $ & $ 2.468(49) $ & $ 0.2147(28) $ & $ 2.221(50) $ \\
$ 6 $ & $ 15.9 $ & $ 0.2313(30) $ & $ 2.317(53) $ & $ 0.2208(29) $ & $ 2.103(47) $ \\
$ 6 $ & $ 16.1 $ & $ 0.2230(38) $ & $ 2.310(56) $ & $ 0.2136(37) $ & $ 2.126(53) $ \\
$ 6 $ & $ 16.3 $ & $ 0.2254(78) $ & $ 2.137(56) $ & $ 0.2162(75) $ & $ 1.975(60) $ \\
\hline
$ 8 $ & $ 26.5 $ & $ 0.2285(31) $ & $ 2.380(49) $ & $ 0.2220(30) $ & $ 2.248(45) $ \\
$ 8 $ & $ 26.7 $ & $ 0.2301(63) $ & $ 2.345(70) $ & $ 0.2243(61) $ & $ 2.230(65) $ \\
$ 8 $ & $ 27.0 $ & $ 0.2285(33) $ & $ 2.351(76) $ & $ 0.2236(32) $ & $ 2.244(66) $ \\
$ 8 $ & $ 27.2 $ & $ 0.2204(32) $ & $ 2.149(64) $ & $ 0.2161(32) $ & $ 2.057(51) $ \\

\hline
\end{tabular}

\end{table}

The topological susceptibility in lattice units $\chi_L(t) a^4$ was
computed for each ensemble separately, using Eq.~(\ref{eq:flowed_topo_suscept}).
The effect of the
rounding procedure, Eq.~(\ref{eq:lat_top_charge_TOT}), on the topological susceptibility, is displayed
in Figure~\ref{fig:TC_vs_t}, where $\chi_L(t)a^4$ is plotted as a function of the flow
time $t/a^2$. For sufficiently large values of the flow time, $\chi_L(t) a^4$
is compatible with a constant within errors. The value of
$\chi_L t_0^2$ computed at the scale $t_0$, obtained
from $c_e=w_e=0.225$, is displayed as a red bullet.
We 
report the results for the choices $c_e=c_w=0.225$ and $c_e=c_w=0.5$,
 in Tables~\ref{tab:top_suscept_0.225_t} and~\ref{tab:top_suscept_0.5_t}, 
respectively. For convenience, we also report the values of 
$\sigma t_0$ and $\sigma w_0^2$, which are taken from Ref.~\cite{Bennett:2020qtj}, except for $\beta=2.55$, $2.65$ for $N_c=2$, and $\beta=7.72$, $7.76$, $7.78$ and $7.80$ for $N_c=4$, which we computed anew.

\begin{table}
\centering
\caption{Continuum limit extrapolations
obtained from the best fit of Eq.~(\ref{eq:contlim})
for the topological susceptibility in $Sp(N_c)$ 
Yang-Mills theories with $N_c=2,\,4,\,6,\,8$.
We present four alternative ways of setting the scale:
we measure $\chi t_0^2(a=0)$ (top section of the table)
and $\chi w_0^4(a=0)$ (bottom),
and adopt as  reference values
either $c_e=0.225=c_w$ (left section of the table)
or $c_e=0.5=c_w$ (right). 
The best-fit curves are displayed as dashed lines in Figure~\ref{fig:contlim}, 
for each value of $N_c$, along 
with the individual measurements. \label{tab:best_fits_contlim}\\}
\begin{tabular}{|c|cc|cc|}
\hline
\hline
$N_c$ & $\chi_L t_0^2(a=0)$ & $\tilde{\mathcal{X}}^2$ & $\chi_L t_0^2(a=0)$ & $\tilde{\mathcal{X}}^2$ \\
 & $c_e = 0.225$ & & $c_e = 0.5$ & \\
\hline
$2$ & $0.000600(22)$ & $1.47$ & $0.002353(93)$ & $1.29$ \\
$4$ & $0.000452(17)$ & $2.18$ & $0.002305(90)$ & $1.89$ \\
$6$ & $0.000315(23)$ & $1.08$ & $0.00185(11)$ & $1.02$ \\
$8$ & $0.000303(23)$ & $2.32$ & $0.00194(15)$ & $1.39$ \\
\hline
\hline
$N_c$ & $\chi_L w_0^4(a=0)$ & $\tilde{\mathcal{X}}^2$ & $\chi_L w_0^4(a=0)$ & $\tilde{\mathcal{X}}^2$ \\
 & $c_w = 0.225$ & & $c_w = 0.5$ & \\
\hline
$2$ & $0.000572(24)$ & $1.16$ & $0.001698(69)$ & $1.50$ \\
$4$ & $0.000584(22)$ & $1.63$ & $0.001991(69)$ & $2.14$ \\
$6$ & $0.000503(32)$ & $1.39$ & $0.00180(12)$ & $1.13$ \\
$8$ & $0.000535(39)$ & $1.41$ & $0.00189(13)$ & $1.80$ \\
\hline
\end{tabular}
\end{table}

Our final results are the extrapolations towards the continuum limit of $\chi_L t_0^2$
for each of the $Sp(N_c)$ Yang-Mills theories.
We obtain these by assuming the following functional dependence:
\beq\label{eq:contlim}
\chi_L(a) t_0^2 = \chi_L(a=0) t_0^2 + c_1 \frac{a^2}{t_0}\,,
\eeq
for each value of $N_c$, separately. 
The results of this extrapolation are reported in Table~\ref{tab:best_fits_contlim} 
(and displayed in Fig.~\ref{fig:contlim}) for $c_e=0.225$ and $c_e=0.5$, respectively.
 For completeness, we also report in the same Table~\ref{tab:best_fits_contlim} 
and Fig.~\ref{fig:contlim} also the extrapolation of $\chi_L w_0^2$, obtained with a similar 
fitting function which includes one correction term ${\cal O}\left(\frac{a^2}{w_0^2}\right)$.
The uncertainty on the extrapolated values are obtained from the maximum likelihood
analysis by including statistical uncertainties only.
 Ref.~\cite{Ce:2016awn}, which reports $\chi t_0^2$ for $SU(N_c)$ with $N_c=3,\, 4,\, 5,\, 6$,
 present the results in units of $t_0$, and for the choice  $c_e=0.225$.
 Unfortunately, a direct comparison is not possible, as the authors of  Ref.~\cite{Ce:2016awn}
 did not report on the common $SU(2)\sim Sp(2)$ case. We will return to the comparison of the
 topological susceptibility in $Sp(N_c)$ and $SU(N_c)$ theories
 in a companion publication~\cite{Bennett:2022gdz}.

\begin{figure*}[t]
\centering
\includegraphics{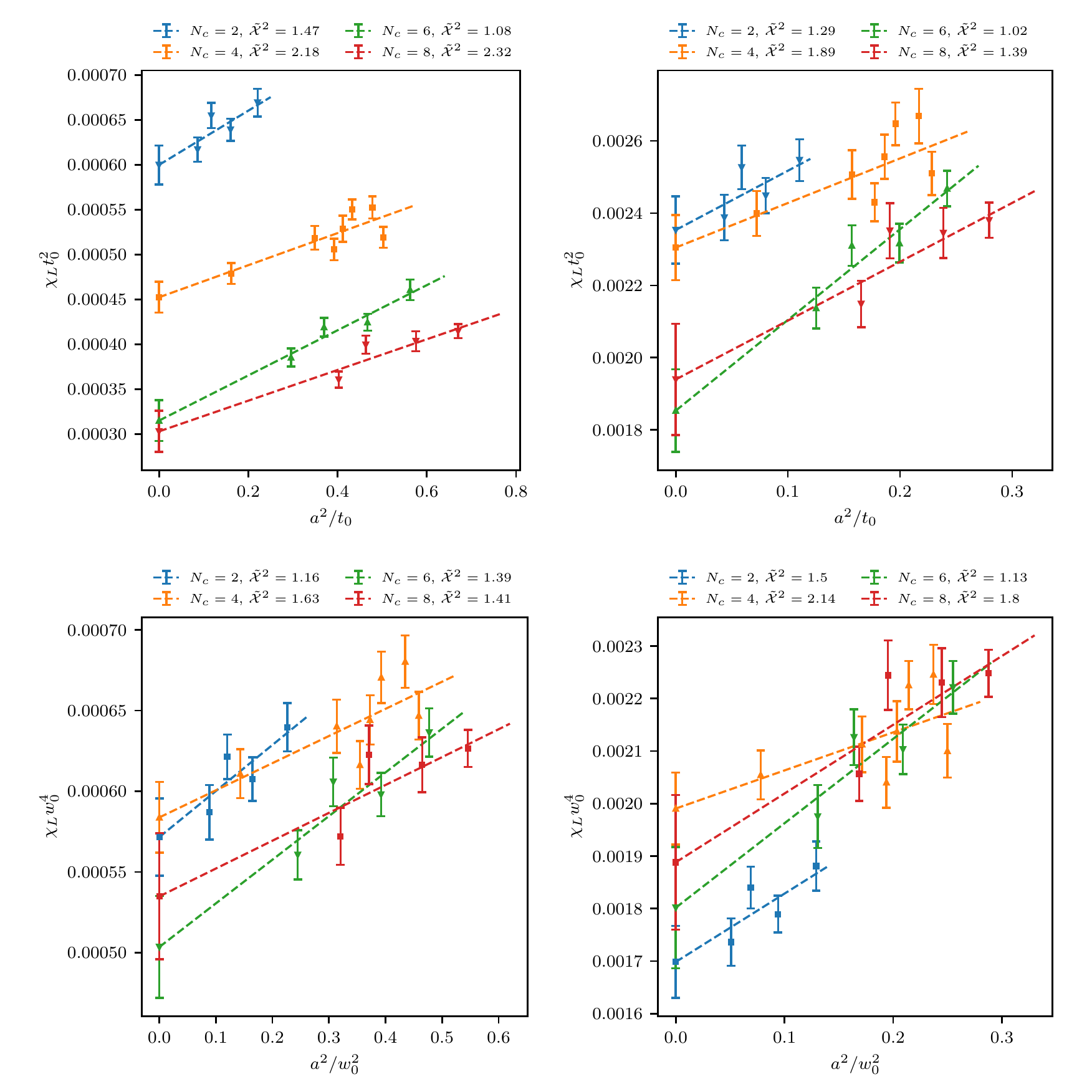}
%\begin{picture}(460,460)
%\put(0,230){\includegraphics[width=.90\columnwidth]{SPN_Topology_contlim_0.225_0.225_scaled.pdf}}
%\put(230,230){\includegraphics[width=.90\columnwidth]{SPN_Topology_contlim_0.5_0.5_scaled.pdf}}
%\put(0,0){\includegraphics[width=.90\columnwidth]{SPN_Topology_contlim_0.225_0.225_scaled_w0.pdf}}
%\put(230,0){\includegraphics[width=.90\columnwidth]{SPN_Topology_contlim_0.5_0.5_scaled_w0.pdf}}
%\end{picture}
\caption{Topological susceptibility per unit volume $\chi_L t_0^2$ as a 
function of $a^2/t_0$ (top panels) and $\chi_L w_0^4$ as a 
function of $a^2/w_0^2$ (bottom), in $Sp(N_c)$ Yang-Mills theories 
with $N_c=2,\,4,\,6,\,8$. We adopt reference values  $c_e=c_w=0.225$ (left panels) and
$c_e=c_w=0.5$ (right).
Our continuum
extrapolations  are represented as 
dashed lines. The results are reported
in Table~\ref{tab:best_fits_contlim}.
\label{fig:contlim}}
\end{figure*}

%%%%%%%%%%%%%%%%%%%%%
%%%%%%%%%%%%%%%%%%%%%
\section{Summary and discussion}
\label{Sec:outlook}

We studied the four-dimensional Yang-Mills theories
with group $Sp(N_c)$, for $N_c=2,\,4,\,6,\,8$, by means of lattice numerical techniques.
We used the Wilson flow as a scale-setting procedure. We showed that
 lattice artefacts are reduced by
adopting the clover-leaf plaquette discretisation. We  also found that by simultaneously rescaling 
 $\mathcal{E}(t)$ (or alternatively $\mathcal{W}(t)$), 
the reference value of  $\mathcal{E}_0$ (or alternatively $\mathcal{W}_0$), 
and $t_0/a^2$ (or alternatively $w_0/a$), the Wilson flows for different 
gauge groups agree (within numerical errors) over a non-trivial range of $t$.
The proposed rescaling is based upon the group-factor dependence of the 
leading-order perturbative  evaluation of $\mathcal{E}(t)$, indicating  that 
the suppression of sub-leading  corrections that 
 differentiate the  groups holds also non-perturbatively,
 over a finite range of flow time $t$.
 
 We then computed the topological susceptibility from our lattice
 ensembles, expressed it in units of the Wilson flow scale $t_0$ ($w_0$),
 and performed the continuum extrapolation for the four  gauge groups.
We summarise in Table~\ref{tab:best_fits_contlim} our results
 for the continuum extrapolation of $\chi t_0^2$ and  $\chi w_0^4$
for two different choices of  reference values for $\mathcal{E}_0$ and $\mathcal{W}_0$.
 In a companion paper~\cite{Bennett:2022gdz}, we present our extrapolation of the measurement of the topological 
 susceptibility towards the large-$N_c$ limit
 (in units of the string tension),  
 and discuss how to compare it with the analogous calculations
 performed for $SU(N_c)$ Yang-Mills theories.

%%%%%%%%%%%
%%%%%%%%%%%
%%%%%%%%%%%

%%%%%%%%%%%%%%%%%%%%%%%%%%%%%%%%%%%%%%%%
%\vspace{1.0cm}

%\FloatBarrier
\begin{acknowledgments}

\vspace{2.0cm}

The work of EB has been funded in part by the Supercomputing Wales project, 
which is part-funded by the European Regional Development Fund (ERDF) via Welsh Government,
and by the UKRI Science and Technology Facilities Council (STFC)
 Research Software Engineering Fellowship EP/V052489/1

The work of DKH was supported by the National Research Foundation of Korea (NRF) grant funded by the Korea government (MSIT) (2021R1A4A5031460) and also by Basic Science Research Program through the National Research Foundation of Korea (NRF) funded by the Ministry of Education (NRF-2017R1D1A1B06033701).

The work of JWL is supported by the National Research Foundation of Korea (NRF) grant funded 
by the Korea government(MSIT) (NRF-2018R1C1B3001379). 

The work of CJDL is supported by the Taiwanese MoST grant 109-2112-M-009-006-MY3. 

The work of BL and MP has been supported in part by the STFC 
Consolidated Grants No. ST/P00055X/1 and No. ST/T000813/1.
 BL and MP received funding from
the European Research Council (ERC) under the European
Union’s Horizon 2020 research and innovation program
under Grant Agreement No.~813942. 
The work of BL is further supported in part 
by the Royal Society Wolfson Research Merit Award 
WM170010 and by the Leverhulme Trust Research Fellowship No. RF-2020-4619.

The work of DV is supported in part by the INFN HPC-HTC project 
and in part by the Simons Foundation under the program “Targeted 
Grants to Institutes” awarded to 
the Hamilton Mathematics Institute.

Numerical simulations have been performed on the 
Swansea University SUNBIRD cluster (part of the Supercomputing Wales project) and AccelerateAI A100 GPU system,
on the local HPC
clusters in Pusan National University (PNU) and in National Yang Ming Chiao Tung University (NYCU),
and the DiRAC Data Intensive service at Leicester.
The Swansea University SUNBIRD system and AccelerateAI are part funded by the European Regional Development Fund (ERDF) via Welsh Government.
The DiRAC Data Intensive service at Leicester is operated by 
the University of Leicester IT Services, which forms part of 
the STFC DiRAC HPC Facility (www.dirac.ac.uk). The DiRAC 
Data Intensive service equipment at Leicester was funded 
by BEIS capital funding via STFC capital grants ST/K000373/1 
and ST/R002363/1 and STFC DiRAC Operations grant ST/R001014/1. 
DiRAC is part of the National e-Infrastructure.

\vspace{1.0cm}

{\bf Open Access Statement - } For the purpose of open access, the authors have applied a Creative Commons 
Attribution (CC BY) licence  to any Author Accepted Manuscript version arising.

\end{acknowledgments}
%%%%%%%%%%%%%%%
%%%%%%%%%%%%%%% Appendix
%%%%%%%%%%%%%%%
%\FloatBarrier
\appendix

\section{Scale setting data for $N_c=2$, $4$, $8$}
\label{Appendix}

In this Appendix, we report the intermediate results of the scale-setting procedure for the $Sp(2)$, $Sp(4)$,
and $Sp(8)$ Yang-Mills theories. The presentation mirrors the one for the $Sp(6)$ theory, in the main body of the paper.
Tables~\ref{tab:scale_2_t0} and~\ref{tab:scale_2_w0} present our results for $t_0/a^2$ and $w_0/a$, respectively,
 for various choices of 
reference values $\mathcal{E}_0$ and $\mathcal{W}_0$, for $N_c=2$.
We tabulate the results for both the plaquette and clover discretisations.
Tables~\ref{tab:scale_4_t0} and~\ref{tab:scale_4_w0} list the same information, but for $Sp(4)$,
while Tables~\ref{tab:scale_8_t0} and~\ref{tab:scale_8_w0} refer to $Sp(8)$.
Cases in which Eqs.~(\ref{eq:scale_t0}) or~(\ref{eq:scale_w0}) do not admit a solution,
because of extreme choices of ${\cal E}_0$ or ${\cal W}_0$, are left blank.
The information in Tables~\ref{tab:scale_2_t0}-\ref{tab:scale_8_w0} 
is also graphically displayed in Figs.~\ref{fig:scale_2_t0}--\ref{fig:scale_8_w0}, and available in machine-readable form in Ref.~\cite{datapackage}.

\begin{figure}[t]
\centering
\includegraphics{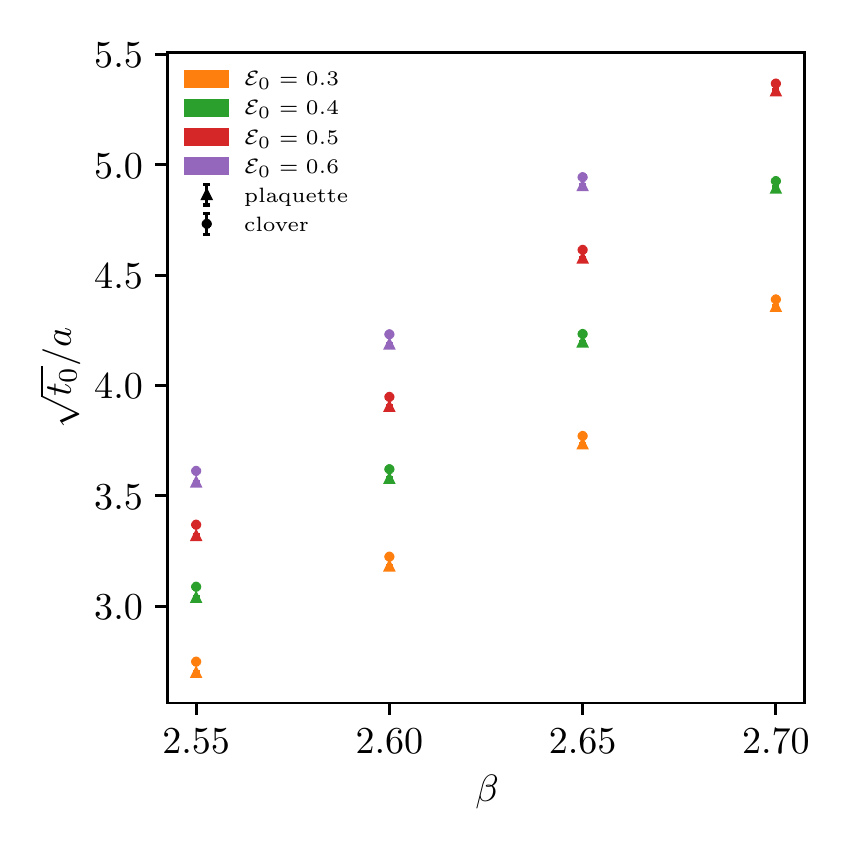}
\caption{The gradient flow scale $\sqrt{t_0}/a$ in $Sp(2)$, 
for different choices of $\mathcal{E}_0$, 
as a function of $\beta$, and comparing plaquette and clover discretisation.\label{fig:scale_2_t0}}
\end{figure}

\begin{figure}[t]
\centering
\includegraphics{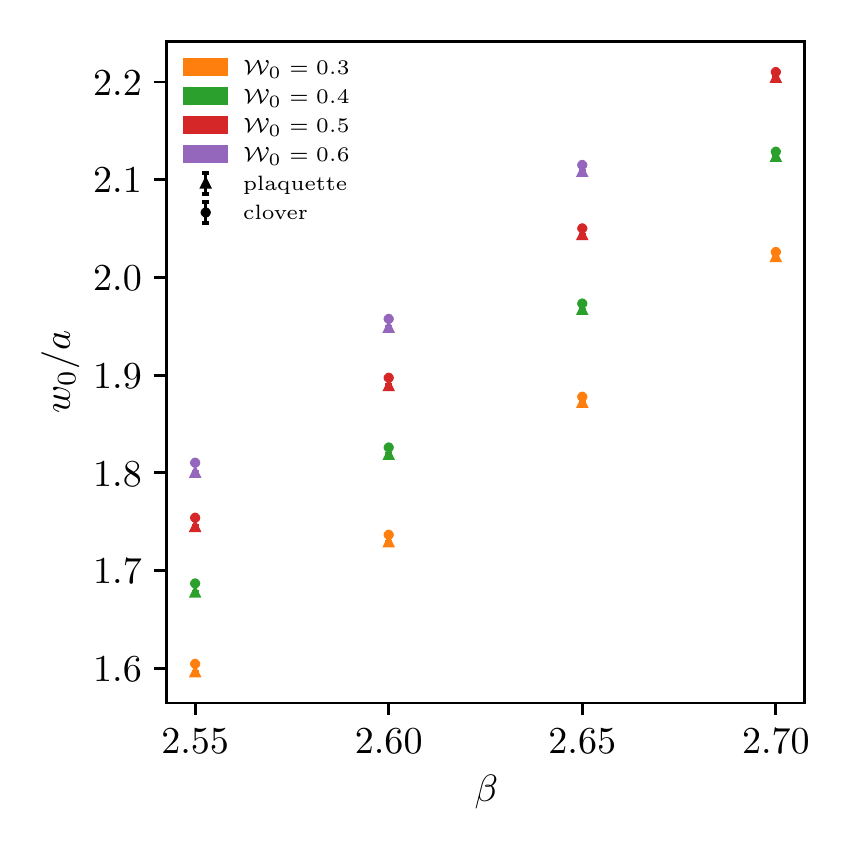}
\caption{The gradient flow scale ${w_0}/a$ in $Sp(2)$, 
for different choices of $\mathcal{W}_0$, 
as a function of $\beta$, and comparing plaquette and clover discretisation.\label{fig:scale_2_w0}}
\end{figure}

\begin{table}
\caption{The gradient flow scale $\sqrt{t_0}/a$ for different choices of $\mathcal{E}_0$ 
and $\beta$, for $N_c=2$. We report the results for both the plaquette and clover discretisations. 
\label{tab:scale_2_t0}\\}
\centering
%\begin{tabular}{ | c|  c | c | c |}
%\hline\hline
%$~~~\beta~~~$ & $~~~\mathcal{E}_0~~~$ & $~~~\sqrt{t_0}/a$\,\,(pl.) $~~~$& $~~~\sqrt{t_0}/a$\,\,(cl.) $~~~$\\
%\hline
%\input{tables/Scale_2_t0.dat}\\
%\hline
%\end{tabular}
\setlength{\tabcolsep}{10pt}
\begin{tabular}{|c|c|c|c|}
\hline
\hline
$\beta$ & $\mathcal{E}_0$ & $\sqrt{t_0}/a$ (pl.) & $\sqrt{t_0}/a$ (cl.) \\
\hline
$ 2.55 $&$ 0.3 $&$ 2.6989(30) $&$ 2.7489(29) $ \\
$ 2.60 $&$ 0.3 $&$ 3.1810(33) $&$ 3.2244(32) $ \\
$ 2.65 $&$ 0.3 $&$ 3.7340(29) $&$ 3.7714(30) $ \\
$ 2.70 $&$ 0.3 $&$ 4.3570(48) $&$ 4.3900(46) $ \\
$ 2.55 $&$ 0.4 $&$ 3.0384(36) $&$ 3.0883(34) $ \\
$ 2.60 $&$ 0.4 $&$ 3.5774(40) $&$ 3.6209(38) $ \\
$ 2.65 $&$ 0.4 $&$ 4.1958(36) $&$ 4.2335(36) $ \\
$ 2.70 $&$ 0.4 $&$ 4.8929(57) $&$ 4.9263(55) $ \\
$ 2.55 $&$ 0.5 $&$ 3.3189(41) $&$ 3.3694(40) $ \\
$ 2.60 $&$ 0.5 $&$ 3.9040(46) $&$ 3.9484(43) $ \\
$ 2.65 $&$ 0.5 $&$ 4.5761(41) $&$ 4.6146(41) $ \\
$ 2.70 $&$ 0.5 $&$ 5.3333(66) $&$ 5.3680(64) $ \\
$ 2.55 $&$ 0.6 $&$ 3.5614(47) $&$ 3.6131(45) $ \\
$ 2.60 $&$ 0.6 $&$ 4.1864(51) $&$ 4.2320(48) $ \\
$ 2.65 $&$ 0.6 $&$ 4.9040(46) $&$ 4.9438(46) $ \\
$ 2.70 $&$ 0.6 $&$ - $&$ - $ \\
\hline
\end{tabular}
\end{table}

\begin{table}
\caption{The gradient flow scale $w_0/a$ for different choices of $\mathcal{W}_0$ 
as a function of $\beta$, for $N_c=2$. We report the results for both the plaquette and clover discretisations. \label{tab:scale_2_w0}\\}
\centering
%\begin{tabular}{ | c|  c | c | c |}
%\hline\hline
%$~~~\beta~~~$ & $~~~\mathcal{W}_0~~~$ & $~~~w_0/a$\,\,(pl.) $~~~$& $~~~w_0/a$\,\,(cl.) $~~~$\\
%\hline
%\input{tables/Scale_2_w0.dat}\\
%\hline
%\end{tabular}
\setlength{\tabcolsep}{10pt}
\begin{tabular}{|c|c|c|c|}
\hline
\hline
$\beta$ & $\mathcal{W}_0$ & $w_0/a$ (pl.) & $w_0/a$ (cl.) \\
\hline
$ 2.55 $&$ 0.3 $&$ 1.5961(11) $&$ 1.6046(10) $ \\
$ 2.60 $&$ 0.3 $&$ 1.7292(11) $&$ 1.7366(10) $ \\
$ 2.65 $&$ 0.3 $&$ 1.87188(92) $&$ 1.87785(90) $ \\
$ 2.70 $&$ 0.3 $&$ 2.0210(14) $&$ 2.0260(13) $ \\
$ 2.55 $&$ 0.4 $&$ 1.6776(12) $&$ 1.6868(11) $ \\
$ 2.60 $&$ 0.4 $&$ 1.8185(13) $&$ 1.8260(12) $ \\
$ 2.65 $&$ 0.4 $&$ 1.9669(10) $&$ 1.9732(10) $ \\
$ 2.70 $&$ 0.4 $&$ 2.1235(16) $&$ 2.1287(15) $ \\
$ 2.55 $&$ 0.5 $&$ 1.7448(13) $&$ 1.7541(13) $ \\
$ 2.60 $&$ 0.5 $&$ 1.8889(14) $&$ 1.8973(13) $ \\
$ 2.65 $&$ 0.5 $&$ 2.0435(11) $&$ 2.0502(11) $ \\
$ 2.70 $&$ 0.5 $&$ 2.2045(18) $&$ 2.2102(16) $ \\
$ 2.55 $&$ 0.6 $&$ 1.8002(15) $&$ 1.8104(14) $ \\
$ 2.60 $&$ 0.6 $&$ 1.9487(14) $&$ 1.9576(14) $ \\
$ 2.65 $&$ 0.6 $&$ 2.1082(13) $&$ 2.1151(12) $ \\
$ 2.70 $&$ 0.6 $&$ - $&$ - $ \\
\hline
\end{tabular}
\end{table}

\begin{table}
\caption{The gradient flow scale $\sqrt{t_0}/a$ for different choices of $\mathcal{E}_0$ 
and $\beta$, for $N_c=4$. We report the results for both the plaquette and clover discretisations. 
\label{tab:scale_4_t0}\\}
\centering
%\begin{tabular}{ | c|  c | c | c |}
%\hline\hline
%$~~~\beta~~~$ & $~~~\mathcal{E}_0~~~$ & $~~~\sqrt{t_0}/a$\,\,(pl.) $~~~$& $~~~\sqrt{t_0}/a$\,\,(cl.) $~~~$\\
%\hline
%\input{tables/Scale_4_t0.dat}\\
%\hline
%\end{tabular}
\setlength{\tabcolsep}{10pt}
\begin{tabular}{|c|c|c|c|}
\hline
\hline
$\beta$ & $\mathcal{E}_0$ & $\sqrt{t_0}/a$ (pl.) & $\sqrt{t_0}/a$ (cl.) \\
\hline
$ 7.7 $&$ 0.3 $&$ 1.37620(40) $&$ 1.46197(40) $ \\
$ 7.72 $&$ 0.3 $&$ 1.41690(43) $&$ 1.49953(42) $ \\
$ 7.76 $&$ 0.3 $&$ 1.49948(50) $&$ 1.57652(50) $ \\
$ 7.78 $&$ 0.3 $&$ 1.54145(54) $&$ 1.61595(53) $ \\
$ 7.80 $&$ 0.3 $&$ 1.58361(60) $&$ 1.65573(57) $ \\
$ 7.85 $&$ 0.3 $&$ 1.68978(70) $&$ 1.75659(67) $ \\
$ 7.7 $&$ 0.4 $&$ 1.62985(53) $&$ 1.69845(52) $ \\
$ 7.72 $&$ 0.4 $&$ 1.67629(55) $&$ 1.74253(55) $ \\
$ 7.76 $&$ 0.4 $&$ 1.77067(65) $&$ 1.83274(65) $ \\
$ 7.78 $&$ 0.4 $&$ 1.81913(70) $&$ 1.87920(70) $ \\
$ 7.80 $&$ 0.4 $&$ 1.86744(78) $&$ 1.92574(75) $ \\
$ 7.85 $&$ 0.4 $&$ 1.99001(92) $&$ 2.04422(88) $ \\
$ 7.7 $&$ 0.5 $&$ 1.82938(63) $&$ 1.89073(62) $ \\
$ 7.72 $&$ 0.5 $&$ 1.88069(66) $&$ 1.94002(66) $ \\
$ 7.76 $&$ 0.5 $&$ 1.98496(78) $&$ 2.04067(77) $ \\
$ 7.78 $&$ 0.5 $&$ 2.03884(84) $&$ 2.09278(84) $ \\
$ 7.80 $&$ 0.5 $&$ 2.09229(94) $&$ 2.14469(91) $ \\
$ 7.85 $&$ 0.5 $&$ 2.2286(11) $&$ 2.2774(11) $ \\
$ 7.7 $&$ 0.6 $&$ 1.99738(73) $&$ 2.05499(71) $ \\
$ 7.72 $&$ 0.6 $&$ 2.05290(76) $&$ 2.10871(76) $ \\
$ 7.76 $&$ 0.6 $&$ 2.16562(90) $&$ 2.21813(89) $ \\
$ 7.78 $&$ 0.6 $&$ 2.22422(97) $&$ 2.27504(97) $ \\
$ 7.80 $&$ 0.6 $&$ 2.2821(11) $&$ 2.3315(10) $ \\
$ 7.85 $&$ 0.6 $&$ 2.4302(13) $&$ 2.4762(12) $ \\
\hline
\end{tabular}
\end{table}

\begin{table}
\caption{The gradient flow scale $w_0/a$ for different choices of $\mathcal{W}_0$ 
as a function of $\beta$, for $N_c=4$. We report the results for both the plaquette and clover discretisations. \label{tab:scale_4_w0}\\}
\centering
%\begin{tabular}{ | c|  c | c | c |}
%\hline\hline
%$~~~\beta~~~$ & $~~~\mathcal{W}_0~~~$ & $~~~w_0/a$\,\,(pl.) $~~~$& $~~~w_0/a$\,\,(cl.) $~~~$\\
%\hline
%\input{tables/Scale_4_w0.dat}\\
%\hline
%\end{tabular}
\setlength{\tabcolsep}{10pt}
\begin{tabular}{|c|c|c|c|}
\hline
\hline
$\beta$ & $\mathcal{W}_0$ & $w_0/a$ (pl.) & $w_0/a$ (cl.) \\
\hline
$ 7.7 $&$ 0.3 $&$ 1.22784(24) $&$ 1.23049(24) $ \\
$ 7.72 $&$ 0.3 $&$ 1.24453(25) $&$ 1.24707(25) $ \\
$ 7.76 $&$ 0.3 $&$ 1.27680(29) $&$ 1.27933(28) $ \\
$ 7.78 $&$ 0.3 $&$ 1.29372(31) $&$ 1.29613(30) $ \\
$ 7.80 $&$ 0.3 $&$ 1.30975(34) $&$ 1.31216(32) $ \\
$ 7.85 $&$ 0.3 $&$ 1.35073(39) $&$ 1.35299(37) $ \\
$ 7.7 $&$ 0.4 $&$ 1.29494(27) $&$ 1.30037(27) $ \\
$ 7.72 $&$ 0.4 $&$ 1.31252(28) $&$ 1.31777(28) $ \\
$ 7.76 $&$ 0.4 $&$ 1.34674(33) $&$ 1.35169(32) $ \\
$ 7.78 $&$ 0.4 $&$ 1.36470(35) $&$ 1.36940(34) $ \\
$ 7.80 $&$ 0.4 $&$ 1.38167(38) $&$ 1.38628(37) $ \\
$ 7.85 $&$ 0.4 $&$ 1.42523(44) $&$ 1.42944(42) $ \\
$ 7.7 $&$ 0.5 $&$ 1.34954(30) $&$ 1.35640(29) $ \\
$ 7.72 $&$ 0.5 $&$ 1.36784(31) $&$ 1.37449(30) $ \\
$ 7.76 $&$ 0.5 $&$ 1.40345(36) $&$ 1.40970(35) $ \\
$ 7.78 $&$ 0.5 $&$ 1.42221(38) $&$ 1.42816(38) $ \\
$ 7.80 $&$ 0.5 $&$ 1.43994(41) $&$ 1.44572(40) $ \\
$ 7.85 $&$ 0.5 $&$ 1.48552(49) $&$ 1.49076(47) $ \\
$ 7.7 $&$ 0.6 $&$ 1.39593(33) $&$ 1.40365(32) $ \\
$ 7.72 $&$ 0.6 $&$ 1.41485(33) $&$ 1.42232(33) $ \\
$ 7.76 $&$ 0.6 $&$ 1.45159(39) $&$ 1.45864(38) $ \\
$ 7.78 $&$ 0.6 $&$ 1.47110(42) $&$ 1.47776(41) $ \\
$ 7.80 $&$ 0.6 $&$ 1.48942(44) $&$ 1.49585(44) $ \\
$ 7.85 $&$ 0.6 $&$ 1.53659(53) $&$ 1.54244(51) $ \\
\hline
\end{tabular}
\end{table}

\begin{figure}[t]
\centering
\includegraphics{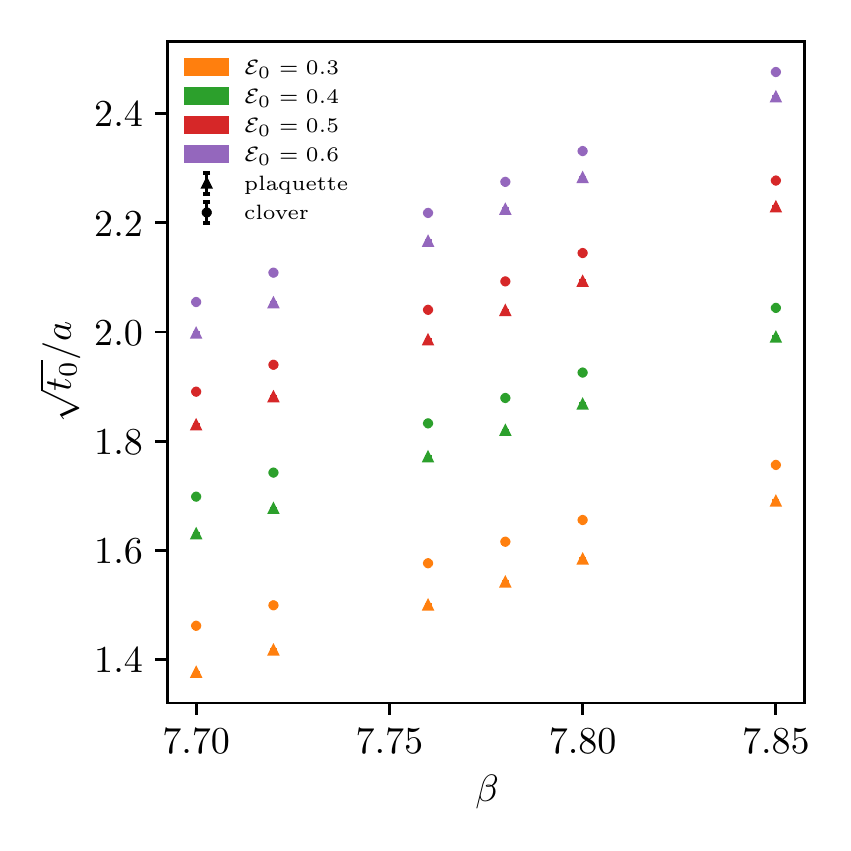}
\caption{The gradient flow scale $\sqrt{t_0}/a$ in $Sp(4)$, 
for different choices of $\mathcal{E}_0$, 
as a function of $\beta$, and comparing plaquette and clover discretisation.\label{fig:scale_4_t0}}
\end{figure}

\begin{figure}[t]
\centering
\includegraphics{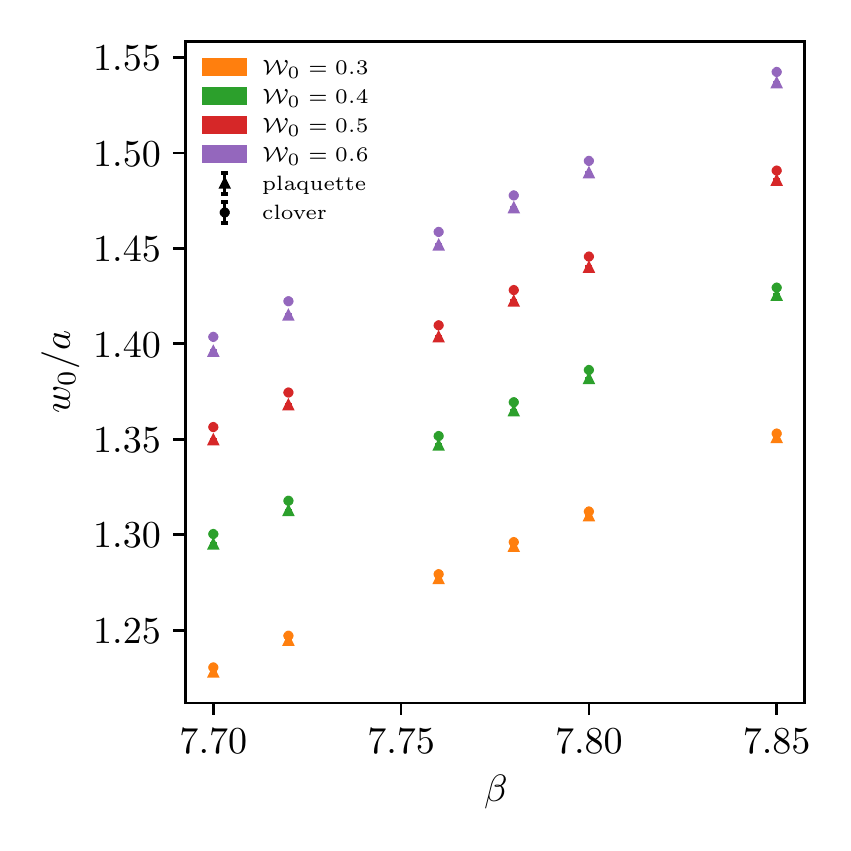}
\caption{The gradient flow scale ${w_0}/a$ in $Sp(4)$, 
for different choices of $\mathcal{W}_0$, 
as a function of $\beta$, and comparing plaquette and clover discretisation.\label{fig:scale_4_w0}}
\end{figure}

\begin{table}
\caption{The gradient flow scale $\sqrt{t_0}/a$ for different choices of $\mathcal{E}_0$ 
and $\beta$, for $N_c=8$. We report the results for both the plaquette and clover discretisations.
 \label{tab:scale_8_t0}\\}
\centering
%\begin{tabular}{ | c|  c | c | c |}
%\hline\hline
%$~~~\beta~~~$ & $~~~\mathcal{E}_0~~~$ & $~~~\sqrt{t_0}/a$\,\,(pl.) $~~~$& $~~~\sqrt{t_0}/a$\,\,(cl.) $~~~$\\
%\hline
%\input{tables/Scale_8_t0.dat}\\
%\hline
%\end{tabular}
\setlength{\tabcolsep}{10pt}
\begin{tabular}{|c|c|c|c|}
\hline
\hline
$\beta$ & $\mathcal{E}_0$ & $\sqrt{t_0}/a$ (pl.) & $\sqrt{t_0}/a$ (cl.) \\
\hline
$ 26.5 $&$ 0.3 $&$ - $&$ - $ \\
$ 26.7 $&$ 0.3 $&$ - $&$ - $ \\
$ 27.0 $&$ 0.3 $&$ - $&$ - $ \\
$ 27.2 $&$ 0.3 $&$ - $&$ - $ \\
$ 26.5 $&$ 0.4 $&$ 0.6428(11) $&$ 1.04083(17) $ \\
$ 26.7 $&$ 0.4 $&$ 0.91955(26) $&$ 1.12079(20) $ \\
$ 27.0 $&$ 0.4 $&$ 1.09768(30) $&$ 1.24650(27) $ \\
$ 27.2 $&$ 0.4 $&$ 1.20530(34) $&$ 1.33540(35) $ \\
$ 26.5 $&$ 0.5 $&$ 1.08015(25) $&$ 1.21190(23) $ \\
$ 26.7 $&$ 0.5 $&$ 1.19344(27) $&$ 1.30764(27) $ \\
$ 27.0 $&$ 0.5 $&$ 1.36119(40) $&$ 1.45737(38) $ \\
$ 27.2 $&$ 0.5 $&$ 1.47602(46) $&$ 1.56329(49) $ \\
$ 26.5 $&$ 0.6 $&$ 1.25801(30) $&$ 1.35615(28) $ \\
$ 26.7 $&$ 0.6 $&$ 1.37719(33) $&$ 1.46475(34) $ \\
$ 27.0 $&$ 0.6 $&$ 1.55864(49) $&$ 1.63417(47) $ \\
$ 27.2 $&$ 0.6 $&$ 1.68496(57) $&$ 1.75412(61) $ \\
\hline
\end{tabular}
\end{table}

\begin{table}
\caption{The gradient flow scale $w_0/a$ for different choices of $\mathcal{W}_0$ 
as a function of $\beta$, for $N_c=8$. We report the results for both the plaquette and clover discretisations. \label{tab:scale_8_w0}\\}
\centering
%\begin{tabular}{ | c|  c | c | c |}
%\hline\hline
%$~~~\beta~~~$ & $~~~\mathcal{W}_0~~~$ & $~~~w_0/a$\,\,(pl.) $~~~$& $~~~w_0/a$\,\,(cl.) $~~~$\\
%\hline
%\input{tables/Scale_8_w0.dat}\\
%\hline
%\end{tabular}
\setlength{\tabcolsep}{10pt}
\begin{tabular}{|c|c|c|c|}
\hline
\hline
$\beta$ & $\mathcal{W}_0$ & $w_0/a$ (pl.) & $w_0/a$ (cl.) \\
\hline
$ 26.5 $&$ 0.3 $&$ - $&$ - $ \\
$ 26.7 $&$ 0.3 $&$ - $&$ - $ \\
$ 27.0 $&$ 0.3 $&$ - $&$ - $ \\
$ 27.2 $&$ 0.3 $&$ - $&$ - $ \\
$ 26.5 $&$ 0.4 $&$ 1.11790(16) $&$ 1.10726(15) $ \\
$ 26.7 $&$ 0.4 $&$ 1.16124(17) $&$ 1.15275(18) $ \\
$ 27.0 $&$ 0.4 $&$ 1.22617(25) $&$ 1.21983(24) $ \\
$ 27.2 $&$ 0.4 $&$ 1.27059(28) $&$ 1.26524(30) $ \\
$ 26.5 $&$ 0.5 $&$ 1.16496(18) $&$ 1.16048(17) $ \\
$ 26.7 $&$ 0.5 $&$ 1.21151(19) $&$ 1.20807(20) $ \\
$ 27.0 $&$ 0.5 $&$ 1.28068(28) $&$ 1.27826(27) $ \\
$ 27.2 $&$ 0.5 $&$ 1.32779(31) $&$ 1.32582(34) $ \\
$ 26.5 $&$ 0.6 $&$ 1.20570(19) $&$ 1.20466(18) $ \\
$ 26.7 $&$ 0.6 $&$ 1.25467(21) $&$ 1.25407(22) $ \\
$ 27.0 $&$ 0.6 $&$ 1.32710(31) $&$ 1.32690(29) $ \\
$ 27.2 $&$ 0.6 $&$ 1.37632(34) $&$ 1.37627(37) $ \\
\hline
\end{tabular}
\end{table}

%%%%%%
%%%%%%

\begin{figure}[t]
\centering
\includegraphics{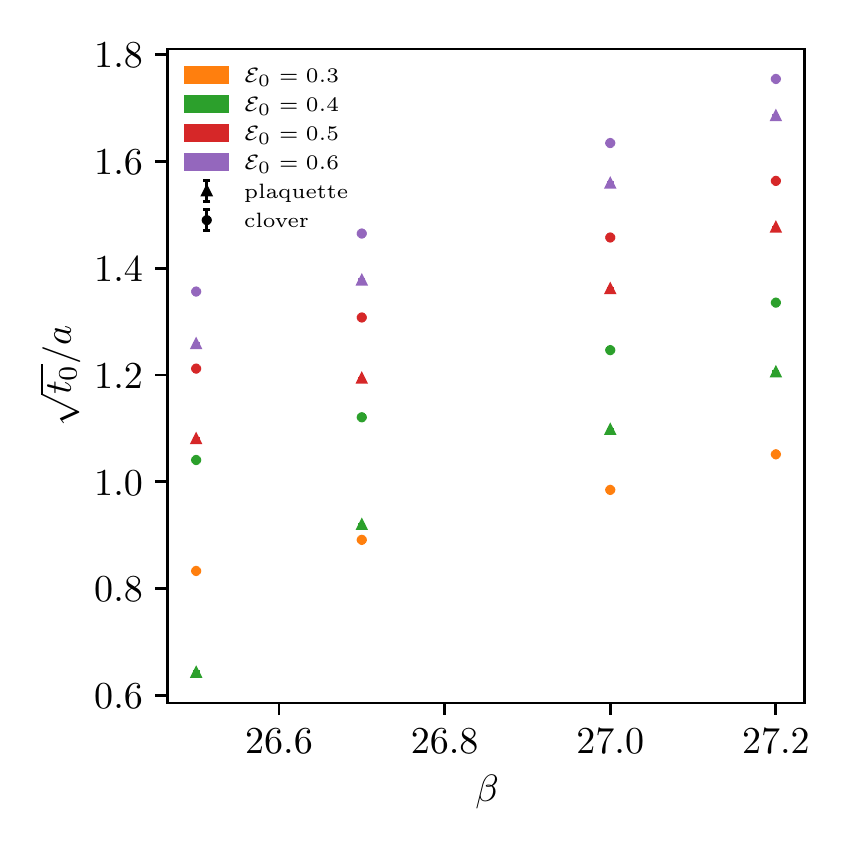}
\caption{The gradient flow scale $\sqrt{t_0}/a$ in $Sp(8)$, 
for different choices of $\mathcal{E}_0$, 
as a function of $\beta$, and comparing plaquette and clover discretisation.\label{fig:scale_8_t0}}
\end{figure}

\begin{figure}[t]
\centering
\includegraphics{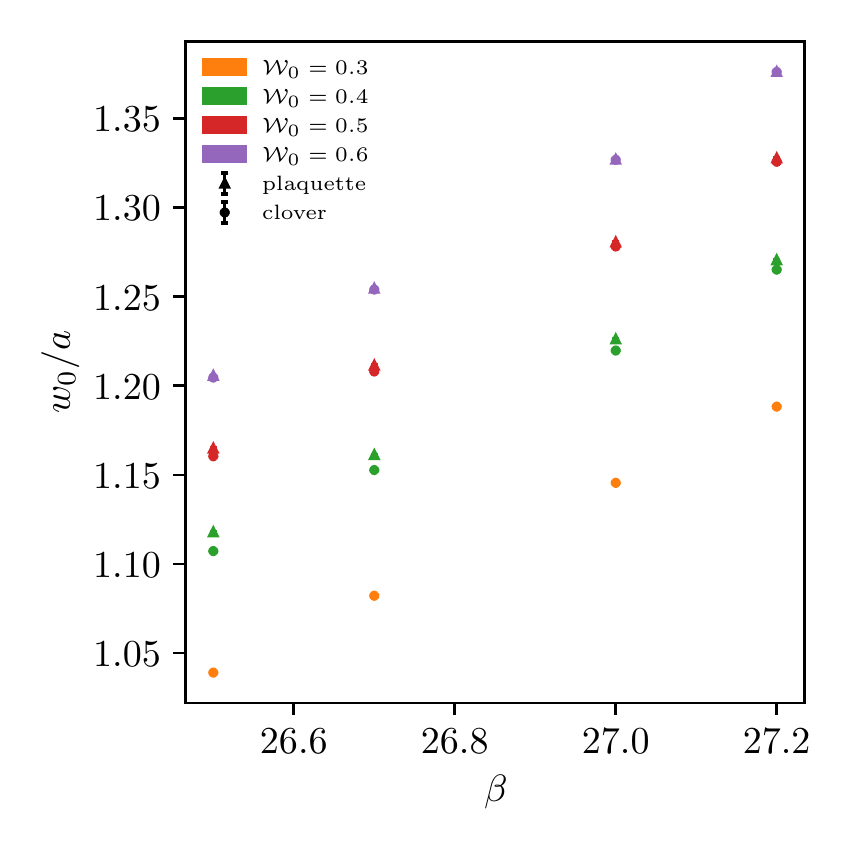}
\caption{The gradient flow scale ${w_0}/a$ in $Sp(8)$, 
for different choices of $\mathcal{W}_0$, 
as a function of $\beta$, and comparing plaquette and clover discretisation.\label{fig:scale_8_w0}}
\end{figure}

%%%%%%%%%%%%
%%%%%%%%%%%% Bibliography
%%%%%%%%%%%%

\end{document}